\documentclass[twocolumn]{aastex701}

\usepackage{graphicx}
\usepackage{amsmath}
\usepackage{hyperref}

\submitjournal{\apj}
\received{XXX}
\revised{YYY}
\accepted{ZZZ}

\begin{document}

\title{The Merger-Driven Formation of Classical Low Surface Brightness Galaxies in {\sc Romulus25}}
\correspondingauthor{Anna C. Wright}
\email{awright@flatironinstitute.org}

\author[0000-0002-1685-5818]{Anna C.\ Wright}
\affiliation{Department of Physics \& Astronomy, Rutgers, The State University of New Jersey, 136 Frelinghuysen Road, Piscataway, NJ 08854, USA}
\affiliation{Center for Astrophysical Sciences, William H.\ Miller III Department of Physics \& Astronomy, Johns Hopkins University, 3400 N.\ Charles Street, Baltimore, MD 21218}
\affiliation{Center for Computational Astrophysics, Flatiron Institute, 162 Fifth Avenue, New York, NY 10010, USA}
\email{awright@flatironinstitute.org}

\author[0000-0002-0372-3736]{Alyson M. Brooks}
\affiliation{Department of Physics \& Astronomy, Rutgers, The State University of New Jersey, 136 Frelinghuysen Road, Piscataway, NJ 08854, USA}
\affiliation{Center for Computational Astrophysics, Flatiron Institute, 162 Fifth Avenue, New York, NY 10010, USA}
\email{abrooks@physics.rutgers.edu}

\author[0000-0002-4353-0306]{Michael Tremmel}
\affiliation{School of Physics, University College Cork, Cork, Ireland}
\email{mtremmel@ucc.ie}

\author[0000-0002-5830-9233]{Jason E. Young}
\affiliation{Department of Astronomy, Williams College, Williamstown, MA 01267, USA}
\affiliation{SETI Institute, 339 Bernardo Ave, Suite 200, Mountain View, CA 94043, United States}
\email{jey1@williams.edu}

\author[0000-0002-9581-0297]{Ferah Munshi}
\affiliation{Department of Physics \& Astronomy, George Mason University, 4400 University Drive, MSN: 3F3, Fairfax, VA 22030-4444, USA}
\email{fmunshi@gmu.edu}

\author[0000-0001-5510-2803]{Thomas R. Quinn}
\affiliation{Astronomy Department, University of Washington, Box 351580, Seattle, WA, 98195-1580}
\email{trq@astro.washington.edu}

\shorttitle{Classical LSB Galaxies}
\shortauthors{A. C. Wright et al.}

\graphicspath{{./}{Figures/}}

\label{firstpage}

\begin{abstract}
We use the \textsc{Romulus25} cosmological simulation volume to study a large sample of late-type gas-rich galaxies with low central surface brightnesses known as classical low surface brightness (LSB) galaxies and compare them to a mass-matched sample of high surface brightness (HSB) galaxies. We find that classical LSB galaxies make up a substantial fraction of the galaxy population, accounting for $\sim$60\% of all central galaxies with 8$\leq$log$_\mathrm{10}$(M$_\star$/M$_\odot$)$\leq$10. In \textsc{Romulus25}, classical LSB galaxies are predominantly formed through major mergers in which the secondary galaxy is co-rotating and aligned with the primary galaxy’s gas disk and/or has above average orbital angular momentum at infall. The merger product is a high spin galaxy in which star formation is spread out and inefficient, allowing the galaxy to build up a large supply of relatively unenriched gas. The star formation rates of LSB galaxies are nearly constant over time, leading to stellar populations that are, on average, slightly older and therefore optically redder than those of similar HSB galaxies. However, because LSB galaxies are diffuse and metal-poor, they have very little internal reddening, causing them to appear bluer than HSB galaxies. We also find that, when compared to the bulges of HSB galaxies, the bulges of LSB galaxies are similar in mass, but are lower surface brightness, redder, and more diffuse on average. Despite these differences, classical LSB galaxies are part of the continuum of the galaxy population in \textsc{Romulus25}, constituting one of many evolutionary paths.
\end{abstract}

\section{Introduction}
\label{intro}
The low surface brightness galaxy family is incredibly diverse, ranging from the faintest dwarf satellites of the Milky Way to some of the largest disk galaxies ever discovered \citep[e.g.,][]{bothun1987,schombert1992,sprayberry1995TFR,greco2018illuminating}. In this paper, we will focus on the late-type population, often referred to as ``classical'' low surface brightness (LSB) galaxies. Although no universal definition of a classical LSB galaxy exists and there is no break in the surface brightness distribution, one of the most broadly adopted identifiers is a B-band central surface brightness ($\mu_\mathrm{0,B}$) fainter than 22.5 mag/arcsec$^2$ \citep[e.g.,][]{deblok1995,morshidiesslinger1999,rosenbaum2004,du2015}. Canonically, these galaxies are characterized by large reservoirs of neutral hydrogen \citep[HI; e.g.,][]{romanishin1982,mcgaugh1994structure,Schombert1997,pildis1997,vanzee1997,burkholder2001,huang2014,du2015,honey2018}, low metallicities and dust contents \citep[e.g.,][]{mcgaugh1994oxygen,ronnback1995,deblok1998oxygen,kuziodenaray2004,rahman2007,Junais2023}, and blue colors \citep[e.g.,][]{mcgaugh1994oxygen,sprayberry1995giant,pildis1997,schombert2014spitzer} that seem contrary to their low past and present star formation rates \citep[e.g.,][]{schombert1992,vanderhulst1993,vandenhoek2000,schombert2011,schombert2015,lei2018}. Taken together, these traits suggest a population that is evolving at a much slower pace than more ``typical'' high surface brightness (HSB) galaxies \citep[e.g.,][]{McGaugh1997,deblok1997,vandenhoek2000}.

The reason for this slow-paced evolution is unknown, but is most commonly attributed to some combination of the low metallicities and low gas surface densities that have been observed in many LSB galaxies \citep[e.g.,][]{vanderhulst1993,deblok1996density,deblok1996HI,vanzee1997,pickering1997,bell2000,leroy2008}. Although they are relatively HI-rich, these galaxies are frequently lacking in molecular hydrogen \citep[H$_2$; e.g.,][]{schombert1990,deblok1998co,oneil2003}, with star formation efficiencies five to ten times lower than those of star-forming HSB galaxies with similar HI reservoirs \citep{boissier2008,wyder2009,lei2018}. Although this has led some to speculate that star formation must proceed differently at low densities \citep[e.g.,][]{weidner2005,helmboldt2005}, observations of resolved stellar populations and individual H~{\sc ii} regions suggest that LSB galaxies form stars via the same underlying mechanisms as HSB galaxies \citep{schombert2011,schombert2013,Young2020UGC8839}.

This, then, raises the question of how LSB galaxies have maintained such a large reservoir of HI without a significant portion of it fragmenting and collapsing into star-forming gas. One possibility is that LSB galaxies form in dark matter halos with higher than average specific angular momentum, resulting in extended low density disks dominated by dark matter at all radii and therefore relatively stable to perturbations \citep[e.g.,][]{dalcanton1997,hernandez1998,mo1998,jimenez1998,jimenez2003,bailin2005,maccio2007,kim2013,cervantessodi2017}. This scenario would also help to explain observations that indicate that LSB galaxies tend to have longer scale lengths and more dark matter governed dynamics than HSB galaxies \citep[e.g.,][]{sprayberry1995TFR,zwaan1995,deblok1996density,deblok1997,verheijen1999,deblok2001mass,swaters2003,Zhong2008,du2019}. Some have also suggested that LSB galaxies preferentially form in low density environments \citep[e.g.,][]{hoffman1992}. Such galaxies would not only be late-forming occupants of low density halos \citep[e.g.,][]{dekel1986,mcgaugh1992,Dominguez-Gomez2023}, they would also be subjected to fewer tidal interactions that might induce instabilities and subsequent bursts of star formation -- a hypothesis that has some observational support \citep[e.g.,][]{zaritsky1993,bothun1993,mo1998,rosenbaum2004,rosenbaum2009}.

The ability of simulations to track the evolution of individual galaxies over time and their lack of overt surface brightness limits make simulations a key part of solving the puzzle of LSB galaxy formation. The earliest simulations of LSB galaxies primarily consisted of N-body simulations, numerical models, and idealized simulations of individual galaxies that examined the stability of LSB disks \citep[e.g.,][]{mihos1997,gerritsen1999,mayer2004}, the influence of specific angular momentum and concentration on galaxy evolution \citep[e.g.,][]{dalcanton1997,mo1998,jimenez1998,bailin2005,maccio2007}, and the viability of various star formation histories as sources of the present day stellar populations and structures of LSB galaxies \citep[e.g.,][]{oneil1998,gerritsen1999,vandenhoek2000,vorobyov2009}. Fully cosmological hydrodynamic simulations, however, have only very recently begun to be used to study LSB galaxies.

Over the last few years, several groups have published results concerning the formation of massive LSB galaxies in cosmological simulations. \citet{zhu2018} found that a Malin 1 analog in TNG100 had been formed when a pair of smaller galaxies fell into the central galaxy, causing formerly hot halo gas to cool and condense into an extended LSB disk. \citet{dicintio2019} used a small suite of zoom-in simulations run with \textsc{NIHAO} to show that aligned gas accretion at high redshift and co-rotating, co-planar mergers tend to produce larger galaxies with lower surface brightnesses than do misaligned interactions. Larger samples from \citet{martin2019} and \citet{Kulier2020} also support the idea that LSB galaxies are formed through external processes. The former found that LSB galaxies in \textsc{Horizon-AGN} have experienced more cumulative tidal perturbation than their HSB counterparts \citep[see also][]{Jackson2021}. However, LSB galaxies in the EAGLE simulations are somewhat isolated (being unlikely to be close-in satellites) and predominantly high-spin, although those with the largest disks also appear to have accreted an abnormally high fraction of their stellar disks from merging galaxies. Higher ex-situ stellar mass fractions combined with higher spin values have also been identified as a common characteristic of LSB galaxies in TNG100 \citep{PerezMontano2022,Zhu2023}. However, it is worth noting that the vast majority of previous simulation studies do not draw a morphological distinction between the classical LSB galaxies described here and the more dominant population of red LSB galaxies that tend to inhabit denser environments.

In this paper, we present results from a large sample of classical LSB galaxies from \textsc{Romulus25}. These galaxies are simulated at higher resolution than the majority of large classical LSB galaxy samples (although see \citeauthor{Jackson2021} \citeyear{Jackson2021}) and are selected via a process that more closely mimics that used by observers. Accordingly, we also endeavor to determine whether the properties of our simulated galaxies match those of observed classical LSB galaxies. We describe our classification process in more detail in Section \ref{lsbclass}. In Section \ref{bary}, we discuss the properties of our sample and compare them to those of observed LSB galaxies. We present our findings on the formation of classical LSB galaxies in Section \ref{form} and summarize our results in Section \ref{lsbsumm}.

\section{The \textsc{Romulus25} Simulation}
\label{sims}
All of the galaxies analyzed in this paper are selected from the high resolution cosmological simulation \textsc{Romulus25} \citep{tremmel2017}. \textsc{Romulus25} is run with the Tree + SPH (smoothed particle hydrodynamics) code \textsc{ChaNGa} \citep{menon2015} and consists of a single (25 Mpc)$^3$ co-moving volume with uniform resolution. The simulation uses a $\Lambda$CDM cosmology following \cite{planck2014} ($\Omega_0$ = 0.3086, $\Lambda$ = 0.6914, $h$ = 0.6777, $\sigma_8$ = 0.8288) and has been evolved until $z=0$. Gravitational interactions between particles are resolved with a spline gravitational force softening length ($\epsilon_g$) of 350 pc (Plummer equivalent 250 pc), which converges to a Newtonian force at 2$\epsilon_g$. \textsc{Romulus25} is initialized with 3.375$\times$ more dark matter particles than gas particles so that similar masses can be used for both: M$_\mathrm{DM,part}$ = 3.39$\times$10$^5$ M$_\odot$ and M$_\mathrm{gas,part}$ = 2.12$\times$10$^5$ M$_\odot$. This reduces numerical heating, which has been shown to lead to unphysical inflation of galaxy sizes \citep{ludlow2019}, and allows the simulation to better track the dynamics of supermassive black holes (SMBHs) \citep{tremmel2015}.

\textsc{ChaNGa} uses many of the same physics modules as its precursor, \textsc{Gasoline} \citep{wadsley2004}, as well as an updated SPH implementation that reduces artificial surface tension by using the geometric mean density in the SPH force expression \citep{Ritchie2001,menon2015,governato2015}, allowing for better capture of fluid instabilities \citep{wadsley2017}. In order to approximate the effects of reionization, \textsc{Romulus25} includes a cosmic UV background \citep{haardt2012} with self-shielding following \citet{pontzen2008}. Gas cooling in low temperature (T $<$ 10$^4$ K) gas is regulated by metal abundance as in \citet{Guedes2011}. However, the simulation does not include high temperature metal-line cooling, which can be a significant cooling mechanism for hot gas in groups and clusters. \citet{christensen2014} shows that the inclusion of high temperature metal-line cooling can contribute to overcooling in simulations like \textsc{Romulus25} that do not have the resolution to model the multi-phase interstellar medium (see \citet{tremmel2019introducing} for an in-depth discussion of this choice). Hot (T $>$ 10$^4$ K) gas therefore cools only via neutral and ionized H and He, with rates calculated from collisional ionization rates \citep{abel1997}, radiative recombination \citep{black1981,verner1996}, photoionization, bremsstrahlung, and H and He line cooling \citep{cen1992}.

Processes that lie below the resolution limit of \textsc{Romulus25}---namely star formation, stellar feedback, black hole growth, and black hole feedback---are handled through subgrid recipes that include empirically chosen parameters. In order to optimize these parameters, \citet{tremmel2017} ran dozens of small zoom-in cosmological simulations with resolution identical to that of \textsc{Romulus25}, varying the values of the free parameters in each. The combination of values that best reproduced the stellar mass--halo mass relation \citep{Moster2013}, the observed gas fraction of HI as a function of stellar mass \citep{Cannon2011,haynes2011}, the mass--spin--morphology relation \citep{Obreschkow2014}, and the SMBH mass--stellar mass relation \citep{schramm2013} were used in the full \textsc{Romulus25} run. The optimization simulations included only isolated galaxies at $z = 0$ with M$_\mathrm{vir}$ = 10$^{10.5-12}$ M$_\odot$, but \textsc{Romulus25} has also been shown to produce realistic galaxies across its entire resolved range (M$_\mathrm{vir}$ = 3$\times$10$^9$ - 2$\times$10$^{13}$ M$_\odot$) and to reproduce observations of high redshift star formation and SMBH growth \citep{tremmel2017}.

In \textsc{Romulus25}, any gas particle that is sufficiently cold ($T$ $<$ 10$^4$ K) and dense ($n$ $>$ 0.2 cm$^{-3}$) has a probability $p$ of forming a star particle:
\begin{equation}
\label{eq:1}
p = \frac{m_{\mathrm{gas}}}{m_{\mathrm{star}}}(1-e^{-c^*\Delta t/t_{\mathrm{dyn}}}),
\end{equation}
where $m_{\mathrm{gas}}$ is the mass of the gas particle, $m_{\mathrm{star}}$ is the mass of the resulting star particle, $c^*$ is the local star-forming efficiency factor (0.15), $\Delta t$ is the timescale over which star formation takes place (10$^6$ yr), and $t_{\mathrm{dyn}}$ is the dynamical time. Each star particle forms with a mass equal to 30\% of the initial gas particle mass (M$_\mathrm{star,part}$ = 6$\times$10$^4$ M$_\odot$) and represents a simple stellar population, with masses, lifetimes, and metal yields drawn from a \citet{kroupa2001} initial mass function. While stars with initial masses greater than 40 M$_\odot$ are assumed to collapse directly to black holes, those with masses between 8 and 40 M$_\odot$ explode as Type II supernovae (SNe), injecting thermal energy into the surrounding interstellar medium (ISM) via `blastwave' feedback \citep{stinson2006} with an energy coupling efficiency of 0.75. Lower mass stars also contribute to feedback via Type Ia SNe and stellar winds. The diffusion of mass, metals, and thermal energy in the ISM follows \citet{shen2010} and \citet{governato2015}.

A gas particle that is eligible to form a star particle may instead form a SMBH if it is sufficiently pristine (metallicity $Z$ $<$ 3$\times$10$^{-4}$), dense ($n$ $>$ 3 cm$^{-3}$), and cold (9 500 $< T$/K $<$ 10 000). These criteria are based on expectations for SMBH formation sites from \citet{Begelman2006} and \citet{Volonteri2012} and ensure that SMBHs form in low metallicity gas that is collapsing faster than the star formation timescale and cooling slowly, while remaining agnostic about the exact formation mechanism. All SMBHs form with a seed mass of 10$^6$ M$_\odot$, accreting mass from surrounding gas particles as necessary, and more than 85\% form within the first Gyr of the simulation \citep{tremmel2018wandering}.

As described in detail in \citet{tremmel2015}, SMBHs can grow by accreting nearby (D $<$ 1.4 kpc) gas or merging with other SMBHs. Gas accretion is governed by a modified Bondi-Hoyle prescription that includes additional angular momentum support based on the resolved dynamics of the gas. Feedback from a given SMBH is proportional to its accretion rate and is isotropically imparted to the surrounding gas particles as thermal energy. 
As in \citet{Bellovary2011}, two SMBHs are considered to have merged when they are within 2 softening lengths of one another and are gravitationally bound. SMBH orbits are traced via the dynamical friction sub-grid model described in \cite{tremmel2015} that produces realistic SMBH sinking and allows the simulation to follow SMBH dynamics even during the mergers of their host galaxies.

We identify halos using Amiga's Halo Finder \citep[AHF;][]{knebe2001,gill2004} and link halos across timesteps with \textsc{tangos} \citep{pontzen2018}. Halo properties are based on all particles within a halo's virial radius (R$_{\mathrm{vir}}$), which is calculated by AHF via a spherical top-hat collapse technique that varies with redshift following \citet{bryan1998}. However, throughout the paper we also refer to M$_{200}$, which is the mass contained within the radius at which the mean enclosed density of particles bound to the halo drops below 200 times the critical density of the universe at the relevant redshift. In order to facilitate better comparison with observations, we calculate stellar masses based on photometric colors following \citet{munshi2013}. However, unless otherwise stated (see Section \ref{col}), we do not attempt to include the effects of dust in our analysis.

\section{Results}
\label{res}
\subsection{Classification of LSB Galaxies}
\label{lsbclass}
In order to facilitate comparisons between observed samples of classical LSB galaxies and our own simulated sample, we attempt to identify candidate galaxies in a way that mirrors observational procedures. Observational samples of classical LSB galaxies are typically selected to be relatively blue and HI-rich (see Sections \ref{HISF} and \ref{col} for discussions of this choice). We therefore limit our analysis to late-type central galaxies, where ``late-type'' galaxies are defined to be those with M$_\mathrm{HI} \geq$ 10$^{6.5}$ M$_\odot$ and B-V $<$ 0.72. These cuts in color and M$_\mathrm{HI}$ are made at values where there is a natural bifurcation in the \textsc{Romulus25} central galaxy sample, but we note that none of our qualitative findings are sensitive to reasonable variations on the locations of these cuts. 

So as to ensure that we are studying only well-resolved galaxies, we also exclude from our sample any galaxies with M$_\mathrm{vir} <$ 3$\times$10$^9$ M$_\odot$ ($\sim$10 000 dark matter particles) or M$_\star <$ 10$^8$ M$_\odot$ ($\sim$1500 star particles), as \citet{wright2021} find that galaxies below this mass show signs of numerical overquenching. Given these criteria, we find a total of 511 late-type central galaxies in \textsc{Romulus25}. In order to calculate the surface brightness of each galaxy, we first compute the B-band luminosity of each star particle based on its mass, metallicity, and age using stellar population synthesis models \citep[http://stev.oapd.inaf.it/cgi-bin/cmd;][]{Marigo2008,Girardi2010}. We rotate each galaxy such that it is face-on and calculate the average B-band surface brightness in radial bins of 300 pc (equal to the spatial resolution of the simulation). We sum the B-band luminosities (in L$_\odot$) of the star particles within a given radial bin, convert this value to absolute magnitudes, and divide by the area of the annulus defined by the bin, converting pc$^2$ to arcsec$^2$ assuming a distance of 10 pc.\footnote{Although this distance is unphysical, surface brightness is independent of distance, if not redshift. Most observational samples of LSB galaxies are, by necessity, at low redshift and therefore only mildly impacted by cosmological dimming, so we do not attempt to correct for it in this paper.} While we use evenly spaced annuli in this analysis, we have confirmed that moderate alterations to the binning scheme (e.g., equal-frequency binning) do not alter our results. We then fit an exponential profile to the resultant B-band surface brightness profile of each galaxy.

The equation for an exponential fit is given by 
\begin{equation}
    \label{eq:diskfit}
    \mu(r) = \mu_0+1.086\frac{r}{r_d},
\end{equation}
where $\mu_0$ is the central surface brightness and $r_\mathrm{d}$ is the scale length of the disk \citep{freeman1970}. The irregular galaxies within our sample are typically well described by a single exponential profile. However, for galaxies with a disk-like morphology, we fit this profile to the disk, avoiding the bulge if one is present via a spatial cut. Disk galaxies are generally defined to be those galaxies with stellar principal axis ratios B/A$\geq$0.75 and C/A$\leq$0.5, although each galaxy is also visually inspected to ensure a sensible morphological classification is made. So as to best mimic observations \citep[e.g.,][]{rosenbaum2009,fathi2010,schombert2011,huang2014,pahwa2018}, we define an LSB galaxy as any galaxy within our sample for which the $y$-intercept of the exponential fit ($\mu_0$) is $\geq$ 22.5 mag/arcsec$^2$. Note that $\mu_0$ should not be interpreted as the actual surface brightness measured at the center of the galaxy, which we denote as $\mu_\mathrm{c}$ here, following \cite{schombert2011}. Due to fluctuations in the radial surface brightness profile and/or the presence of a bulge, $\mu_\mathrm{c}$ could lie above or below $\mu_0$.

From this classification, our sample naturally splits into three categories: irregular LSB galaxies, bulge+disk LSB galaxies (disk galaxies with $\mu_0\geq$ 22.5 mag/arcsec$^2$, but $\mu_\mathrm{c}$ at least 1.25 mag/arcsec$^2$ brighter), and pure disk LSB galaxies (other disk galaxies with $\mu_0\geq$ 22.5 mag/arcsec$^2$). For those galaxies that fall into the bulge+disk category, we also fit a S\'ersic profile to the disk-subtracted surface brightness profile in order to characterize the bulge. The equation for a S\'ersic profile is given by 
\begin{equation}
    \mu(r) = \mu_{\mathrm{eff}}+2.5c_n\Big(\Big(\frac{r}{r_{\mathrm{eff}}}\Big)^{1/n}-1\Big)
\label{eq:ser}
\end{equation}
\citep{sersic1963}, where $\mu_{\mathrm{eff}}$ is the effective surface brightness, $r_{\mathrm{eff}}$ is the effective radius, and c$_n$ = 0.868$n$-0.142, where $n$ is the S\'ersic index \citep{capaccioli1989}. However, note that our nominal definition of a bulge+disk galaxy is extremely simplistic and is intended to identify galaxies with higher surface brightness central regions -- which might therefore have properties and histories distinct from those that are purely low surface brightness -- rather than galaxies that consist of a disk and a classical bulge. Consequently, this subsample may be slightly more heterogeneous than the other two. With this caveat and the general diversity of the galaxy population in mind, we plot the spread in galaxy properties (rather than merely the median behavior) whenever possible throughout the paper.

\begin{figure*}
\centering
\includegraphics[width=0.97\textwidth]{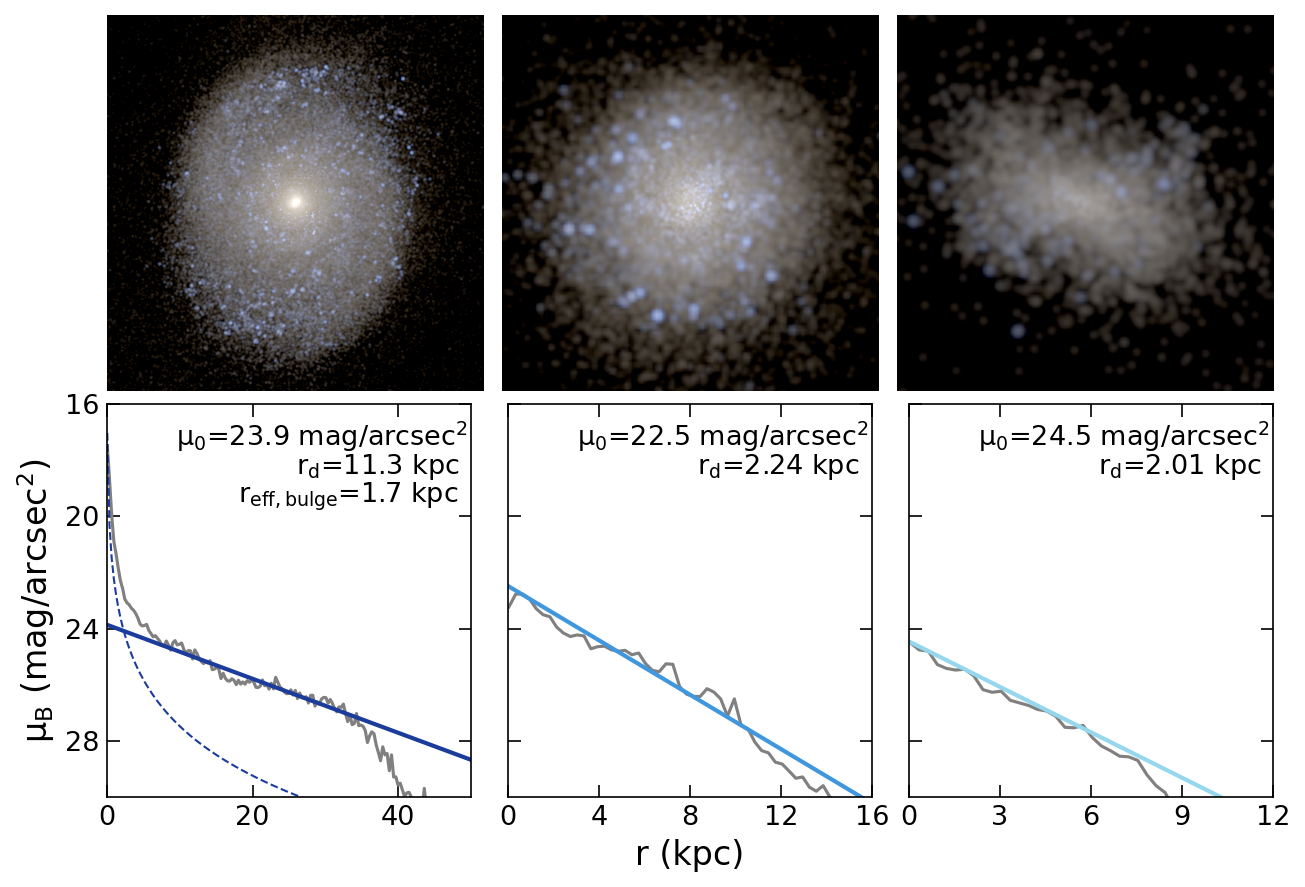}
\caption{B-band surface brightness profiles with accompanying exponential disk (solid blue lines) and S\'ersic bulge (dashed blue line) fits (bottom) and face-on mock UVI images of representative LSB galaxies (top). The top panels are twice the width of the bottom panels and go down to 30 mag/arcsec$^2$. Fit parameters (i.e., central surface brightness, scale length, and, when relevant, bulge effective radius) for each galaxy are shown in the upper right hand corner of each surface brightness profile plot. At left is a bulge+disk LSB galaxy with M$_\star$=4.6$\times$10$^{10}$ M$_\odot$. At center is a pure disk LSB galaxy with M$_\star$=1.6$\times$10$^9$ M$_\odot$. At right is an irregular LSB galaxy with M$_\star$=1.8$\times$10$^8$ M$_\odot$. Surface brightness calculations and UVI images were generated using results from stellar population synthesis models.}
\label{fig:replsb}
\end{figure*}

\begin{figure}
\centering
\includegraphics[width=0.47\textwidth]{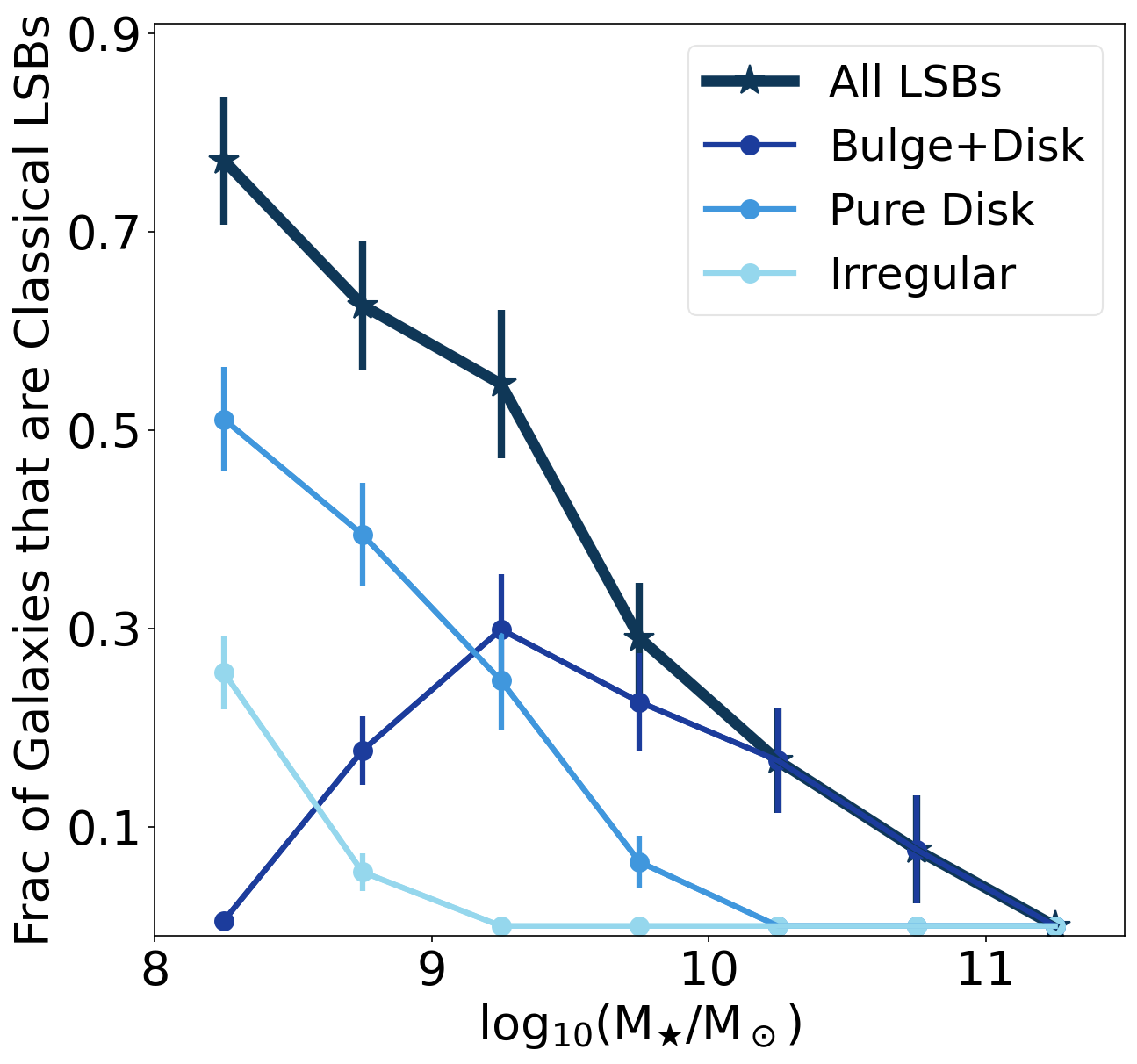}
\caption{The fraction of central {\sc Romulus25} galaxies that are classical LSB galaxies as a function of stellar mass. At higher masses, LSB galaxies are those with fainter central surface brightnesses. At lower masses, nearly all dwarfs have faint central surface brightnesses, and LSB classification is typically more indicative of morphology. Galaxies are grouped in bins of 0.5 dex and errors are Poisson errors.}
\label{fig:frac}
\end{figure}

Of the 326 LSB galaxies in our sample, 55 are irregular galaxies, 89 are bulge+disk galaxies, and 182 are pure disk galaxies. Images and surface brightness profiles of representative examples of each are shown in Figure \ref{fig:replsb}. In Figure \ref{fig:frac}, we plot the fraction of central galaxies that meet our criteria for being classical LSB galaxies as a function of stellar mass. We find that bulge+disk LSB galaxies typically have M$_\star$=10$^{9-11}$ M$_\odot$, pure disk LSB galaxies typically have M$_\star<$10$^{9.5}$ M$_\odot$, and irregular LSB galaxies typically have M$_\star<$10$^{9}$ M$_\odot$. While there is a general trend for bulge+disk LSB galaxies to be the most massive, followed by pure disk LSB galaxies and then irregular LSB galaxies, there is considerable overlap in the stellar masses of our three categories. 

Overall, 60\% of all central galaxies in {\sc Romulus25} with 10$^8<$M$_\star$/M$_\odot<$10$^{10}$ are classical LSB galaxies. The extent to which this value may be in tension with observations is unclear. As early as the 1990s, LSB galaxies were thought to make up a significant fraction of the galaxy population \citep[e.g.,][]{mcgaugh1995,Dalcanton1997Obs}, and more recent large surveys \citep[e.g.,][]{minchin2004,trachternach2006,greco2018illuminating,du2019,Tanoglidis2021} have borne out this idea. However, the 60\% that we find is slightly in excess of even the 50\% quoted in \cite{mcgaugh1995}'s initial estimate. It is certainly possible that \textsc{Romulus25} overproduces LSB galaxies, particularly at the lower end of our mass range. However, most surveys are highly incomplete at the low surface brightnesses---and, in many cases, low luminosities---that characterize these galaxies. Even many surveys that target LSB galaxies would not include LSB disks with HSB bulges as we do here. We may therefore be including many objects that would be undetected in or excluded from observational surveys.

\subsection{Properties of Classical LSB Galaxies}
\label{bary}

\subsubsection{Structural Parameters}
\begin{figure}
\centering
\includegraphics[width=0.29\textwidth]{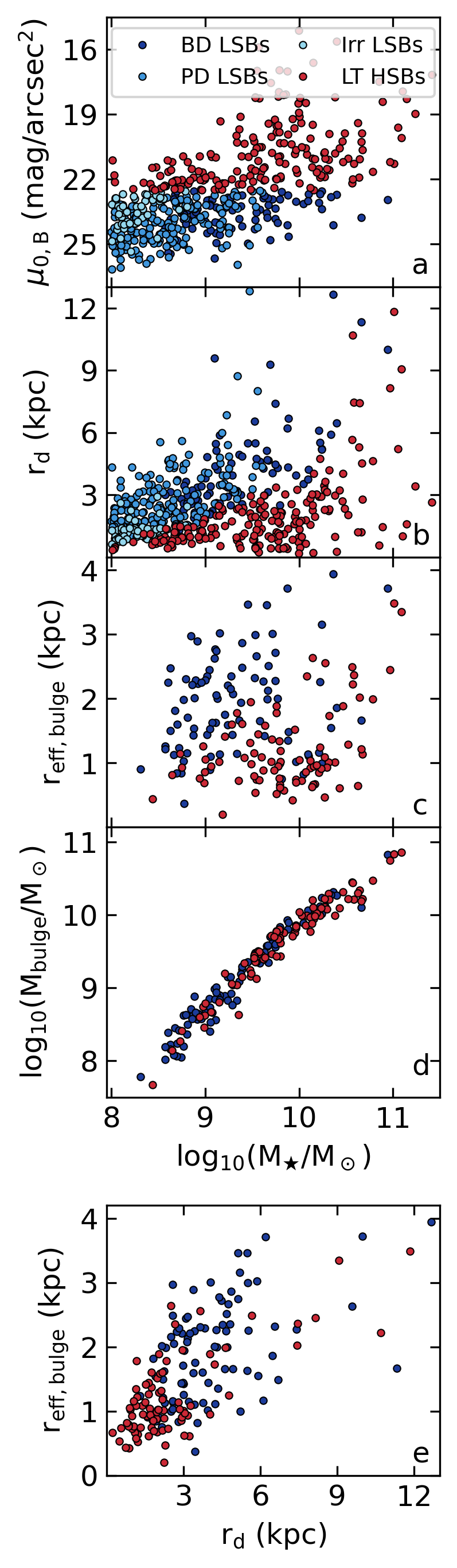}
\caption{The structural parameters of our irregular (Irr), pure disk (PD), and bulge+disk (BD) LSB galaxies (shades of blue) and late-type (LT) HSB galaxies (red). Panels a and b show disk properties (central surface brightness and scale radius), while panels c and d show bulge properties (effective radius and mass) as a function of stellar mass for those galaxies that have bulges. Disk scale radius and bulge effective radius are compared in panel e. LSB galaxies typically have larger disk scale radii and bulge effective radii than HSB galaxies at a given mass. However, the masses of their bulges are typical for their stellar masses and the effective radii of their bulges correlate with the scale radii of their disks.} 
\label{fig:struct}
\end{figure}

In Figure \ref{fig:struct}, we show the structural parameters of the classical LSB and late-type HSB galaxies in \textsc{Romulus25}. Panels a and b show exponential disk fit-based properties ($\mu_\mathrm{0,B}$ and $r_\mathrm{d}$, respectively) for all of the galaxies in the sample, while panels c and d show bulge properties ($r_\mathrm{eff,bulge}$ and $M_\mathrm{bulge}$---defined as the stellar mass contained within $r_\mathrm{eff,bulge}$---respectively) for those galaxies that have bulges according to the criteria discussed in the previous section. As we might expect both from Figure \ref{fig:frac} and from many previous studies \citep[e.g.,][]{Driver1999,Blanton2001}, there is a correlation between stellar mass and $\mu_\mathrm{0,B}$, with HSB galaxies becoming increasingly rare at lower stellar masses. Those HSB galaxies that do exist towards the lower mass end of our sample tend to be unusually compact. By contrast, LSB galaxies are quite extended. Observations have long established that LSB galaxies tend to have high $r_\mathrm{d}$ values \citep[e.g.,][]{sprayberry1995TFR,swaters2003}, but we find that, at a given stellar mass, they also have physically larger bulges than HSB galaxies. This does not, however, imply that their bulges are more massive: as we show in panel d, LSB galaxies and HSB galaxies in \textsc{Romulus25} follow the same tight relation between bulge mass and total stellar mass---possibly a manifestation of the tuning of the simulations to the angular momentum--morphology relation and our inclusion of only galaxies with bright bulges relative to their disks in this subsample. Instead, we see that the bulges of LSB galaxies are, like their disks, more diffuse than those of HSB galaxies. In panel e, we plot the scale radius of the disk against the effective radius of the bulge for these galaxies and show that the two are correlated. As we might expect given Equations \ref{eq:diskfit} and \ref{eq:ser}, we see a similar correlation between $\mu_0$ and $\mu_\mathrm{eff}$. The correlation between r$_\mathrm{d}$ and r$_\mathrm{eff}$ is consistent with observations of both HSB galaxies \citep[e.g.,][]{Courteau1996} and LSB galaxies \citep{Beijersbergen1999,Galaz2006}, suggesting that the formation of the bulge is linked to that of the disk in both populations.

\label{struct}

\subsubsection{HI and Star Formation}
\label{HISF}
\begin{figure*}
\centering
\includegraphics[width=0.97\textwidth]{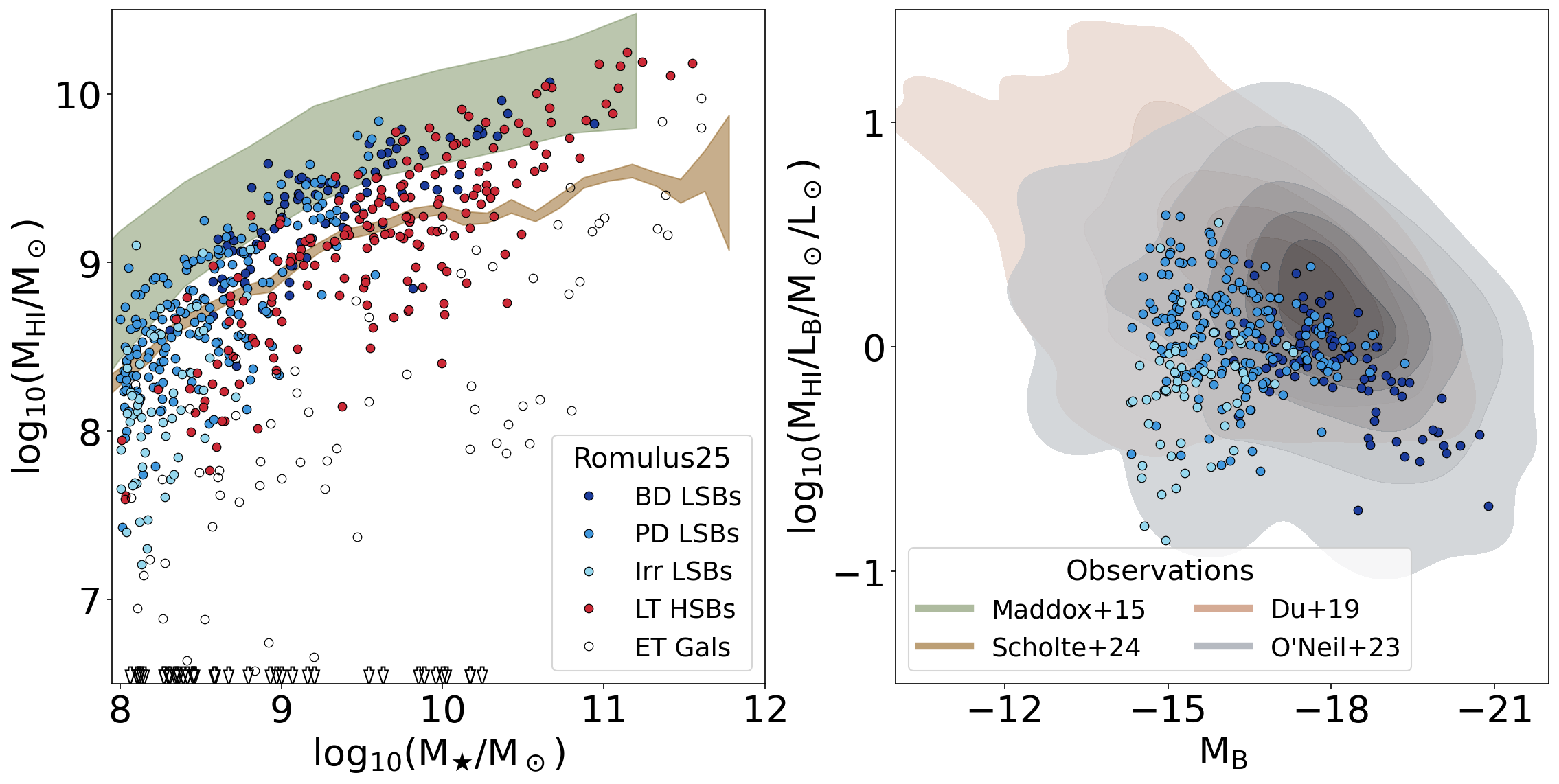}
\caption{\textit{Left:} The stellar and HI masses of LSB galaxies (shades of blue), late-type central HSB galaxies (red), and early-type central galaxies (unfilled) in {\sc Romulus25}. Those galaxies with M$_{\mathrm{HI}}\leq$10$^{6.5}$ M$_\odot$ are indicated with downward-facing arrows. The green band shows the median M$_\star$--M$_\mathrm{HI}$ values from \citet{Maddox2015} with a 1-sigma spread and the gold band shows the median atomic gas sequence from \citet{Scholte2024}. \textit{Right:} The absolute B-band magnitudes and the ratio of HI mass to B-band luminosity for the {\sc Romulus25} LSB galaxies compared to observational samples of LSB galaxies from \citet[][beige]{du2019} and \citet[][gray]{ONeil2023}. Contours indicate the regions containing 95\%, 80\%, 60\%, 40\%, and 20\% of the observational data. The classical LSB galaxies are by far the most HI-rich galaxies in \textsc{Romulus25}. \textsc{Romulus25} does not reproduce the most HI-rich galaxies in the observed LSB galaxy samples, but see the text for a discussion of potential biases.} 
\label{fig:HI}
\end{figure*}

Classical LSB galaxies have been observed to be extremely rich in neutral hydrogen compared to other galaxies \cite[e.g.,][]{mcgaugh1992,deblok1996HI,vandenhoek2000,burkholder2001,oneil2004,du2015,honey2018} and we find that this is also true of our simulated sample. In the left-hand panel of Figure \ref{fig:HI}, we plot the HI and stellar masses of all of the central galaxies in \textsc{Romulus25} with M$_\star\geq$10$^8$ M$\odot$. At any given stellar mass, the LSB galaxies tend to be the most rich in HI. To some extent, this is by design: we have chosen to consider only late-type LSB galaxies in our analysis and contained within that definition is a requirement that these galaxies have M$_{\mathrm{HI}}\geq$10$^{6.5}$ M$_\odot$. However, our LSB galaxies are still HI-rich relative to late-type HSB galaxies, which meet the same color and M$_{\mathrm{HI}}$ criteria. In fact, $\sim$40\% of all neutral hydrogen within {\sc Romulus25} is within LSB galaxies, compared to $\sim$9\% of the simulation's stellar mass.

In the right-hand panel of Figure \ref{fig:HI}, we plot the ratio of M$_\mathrm{HI}$ to B-band luminosity (L$_\mathrm{B}$) against B-band absolute magnitude (M$_\mathrm{B}$) for our LSB galaxies and compare them to two large observational samples of LSB galaxies. The \citet{du2019} sample, which is shown in shades of brown, is made up of over 1000 LSB galaxies identified in the $\alpha$.40-SDSS DR7 survey \citep[see also][]{du2015} and therefore has HI masses from the 40 percent data release of ALFALFA \citep{haynes2011}. The \citet{ONeil2023} sample, depicted in shades of gray, consists of 350 previously-known LSB galaxies for which the authors performed follow-up HI observations (although note that we have only included the 271 galaxies for which the confusion/contamination status was 1 or 2, indicating that the majority of the HI flux is not likely to have been contributed by another galaxy). For each observational sample, the contours shown indicate the regions in the plot containing 95\%, 80\%, 60\%, 40\%, and 20\% of the data. Although both observational samples are weighted towards more massive galaxies than the \textsc{Romulus25} sample (the median M$_\mathrm{B}$ is -16.4 for \textsc{Romulus25} and -16.9 and -17.7 for \citet{du2019} and \citet{ONeil2023}, respectively), we recover the same broad trends: while the lower limit on M$_\mathrm{HI}$/L$_\mathrm{B}$ changes little with the luminosity of the galaxy, the upper limit rises towards fainter/lower mass galaxies, which therefore tend to be the most HI-rich for their stellar masses. However, while the \textsc{Romulus25} LSB galaxies fall within the same range as the observed LSB galaxies on these axes, it is clear that they are not as HI-rich. The median M$_\mathrm{HI}$/L$_\mathrm{B}$ in the \textsc{Romulus25} sample is 0.98 -- much lower than \citet{du2019}'s 2.09 or \citet{ONeil2023}'s 1.31.

There are a number of reasons why the \textsc{Romulus25} LSB galaxies may appear HI-deficient compared to observed LSB galaxies, even though they are the most HI-rich galaxies in the simulation. Perhaps the most significant is that the \textsc{Romulus25} galaxies are overall HI-poor. \citet{Motiwala2025} note that field ultra-diffuse galaxies in both \textsc{Romulus25} and \textsc{NIHAO} are, on average, HI-poor compared to the SMUDGes ultra-diffuse galaxies, and here we show that this discrepancy holds outside of the dwarf regime. While \textsc{Romulus25} reproduces the shape of the \citet{Maddox2015} M$_\star$--M$_\mathrm{HI}$ relation, which combines ALFALFA and SDSS data and is shown as a pale green band in the left-hand panel of Figure \ref{fig:HI}, the simulated galaxies are systematically offset towards low HI masses at all but the highest stellar masses. It is therefore unsurprising that our sample of LSB galaxies fails to reproduce the more HI-rich objects in observed LSB galaxy samples. That being said, HI-selected surveys like ALFALFA tend to be biased high compared to optically-selected surveys \citep[e.g.,][]{catinella2010}. For comparison, we have also included the median atomic gas sequence for the mass-complete sample from \citet{Scholte2024}, which was constructed by stacking the ALFALFA spectra of galaxies from the DESI survey in stellar mass bins of 0.15 dex and is consistent with results from stellar mass-selected surveys like xGASS \citep{Catinella2018}. The gold band representing these data lies below the \citet{Maddox2015} relation at all stellar masses and is considerably closer to what we see in \textsc{Romulus25}. The bias inherent in HI-selected surveys is particularly pervasive in the LSB regime, where redshifts are derived almost exclusively from HI data and galaxies whose HI contents lie below the detection threshold of a given instrument are therefore excluded from catalogs like those we've compared to here \citep[e.g.,][]{oneil2004,ONeil2023}. 

Additionally, the volumes probed by these surveys are much larger than what we simulate in \textsc{Romulus25} and they are therefore far more likely to include rare objects. We do not, for instance, find any LSB galaxies in \textsc{Romulus25} with M$_\mathrm{B}\lesssim$-21 like the brightest/most massive galaxies reported in \citet{ONeil2023}. Giant LSB galaxies, as these are sometimes called, are extremely rare. They are thought to be $\sim$300 times less common than standard L$^\star$ galaxies \citep{briggs1990}, of which \textsc{Romulus25} contains only a handful. Recent estimates from \citet{Saburova2023} based on the local universe ($z<0.1$) place the number density of giant LSB galaxies at 4.04$\times$10$^{-5}$ Mpc$^{-3}$. We would therefore expect to find $<$1 of these extreme galaxies in our small volume. 

\begin{figure*}
\centering
\includegraphics[width=0.97\textwidth]{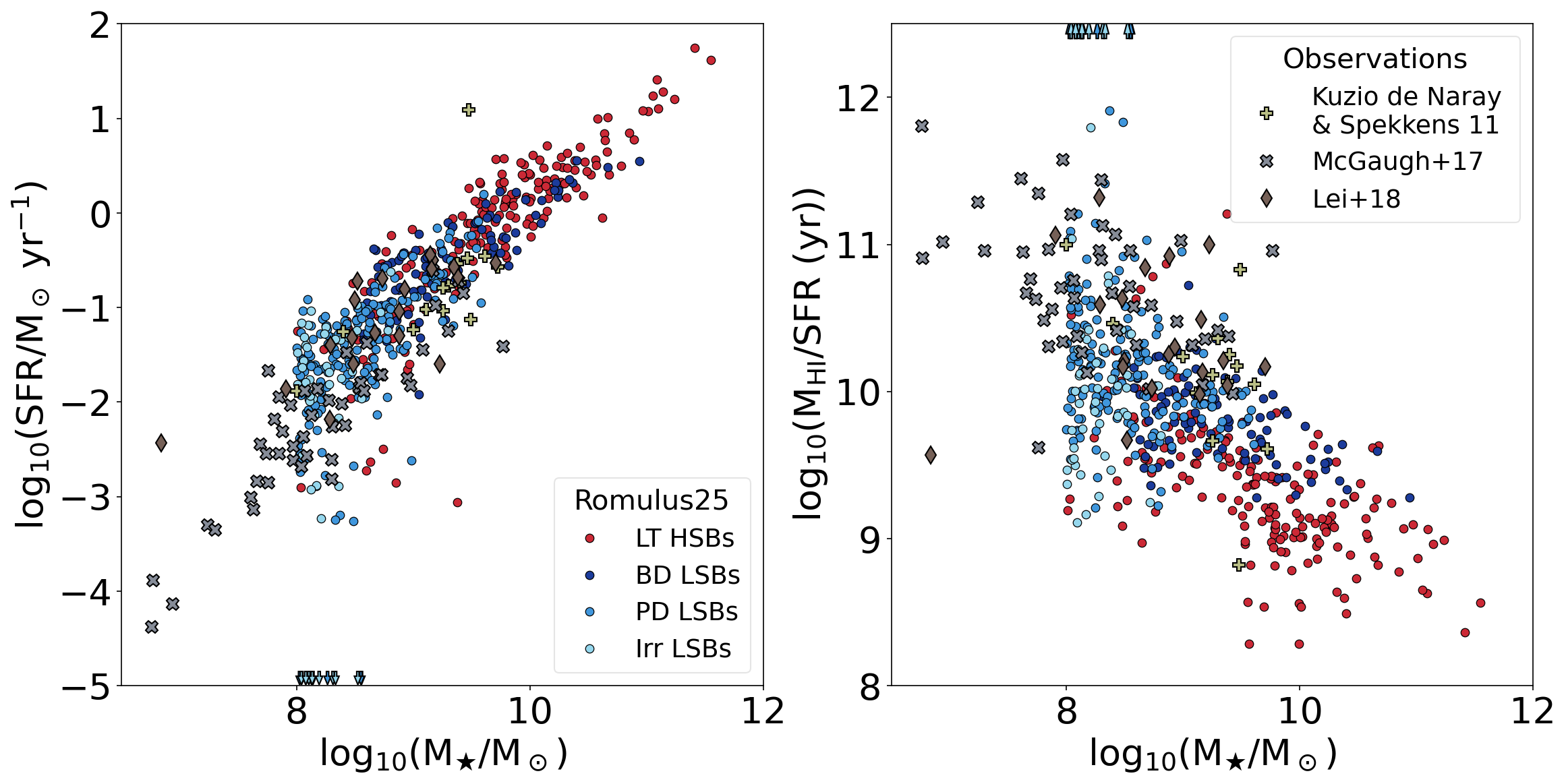}
\caption{\textit{Left:} Instantaneous star formation rates of our classical LSB galaxies (shades of blue) and late-type central HSB galaxies (red) compared to observational data from \protect\cite{kuziodenaray2011} (pale green pluses), \protect\cite{mcgaugh2017} (gray xs), and \protect\cite{lei2018} (brown diamonds). \textsc{Romulus25} galaxies with instantaneous SFR = 0 M$_\odot$/yr are indicated by downward-facing arrows. \textit{Right:} The HI depletion timescales (i.e., the ratio of M$_\mathrm{HI}$ to the instantaneous star formation rate) of the same samples of galaxies. \textsc{Romulus25} galaxies with instantaneous SFRs = 0 M$_\odot$/yr are indicated by upward-facing arrows. At all but the highest masses, the star formation rates of our classical LSB galaxies are typical of late-type galaxies of similar stellar mass. Although this seems contrary to the canonical finding that LSB galaxies have unusually low star formation rates, the star formation rates of our simulated galaxies are largely consistent with those of observed LSB galaxies. Because these galaxies are abnormally HI-rich, it is their HI depletion timescales that are unusual, being, on average, much longer than those of late-type HSB galaxies.}
\label{fig:SFRlsb}
\end{figure*}

Although our simulated LSB galaxies are the most HI-rich galaxies in \textsc{Romulus25} and therefore have an abundance of ostensibly star-forming fuel, they do not have elevated star formation rates relative to other galaxies of the same stellar mass. In the left-hand panel of Figure \ref{fig:SFRlsb}, we compare the instantaneous star formation rates of the LSB galaxies in our sample to those of late-type central HSB galaxies from \textsc{Romulus25} and several observed samples of LSB galaxies. The \citet{lei2018} sample (shown as brown diamonds) is a subset of the \citet{du2019} $\alpha$.40-SDSS DR7 sample for which H$\alpha$-derived star formation rates were acquired; corresponding stellar masses are from the MPA-JHU catalog \citep{Kauffmann2003}. The \citet{mcgaugh2017} galaxies (shown as gray xs) consist of dwarf-mass LSB galaxies, also with H$\alpha$-derived star formation rates. The \citet{kuziodenaray2011} LSB galaxies (shown as pale green pluses) are compiled from a number of sources in the literature; where available, we use the H$\alpha$-derived star formation rates, as the short timescales they probe are more comparable to the instantaneous star formation rates we derive from \textsc{Romulus25}, but there are a few galaxies for which only UV data are available. 

The star formation rates of both our LSB galaxies and our HSB galaxies are largely consistent with those of the observed LSB galaxy samples. However, they are generally biased slightly high with respect to observations---particularly at the low end of our mass range---likely as a result of limited resolution within the simulation. Because \textsc{Romulus25} cannot resolve structures below $\sim$700 pc in size, there is an upper limit on the density that can be resolved. As a result, the star formation prescription implemented in the simulation allows stars to form in relatively underdense gas, which likely boosts star formation rates in galaxies where gas is near the star formation threshold but which might not ordinarily form stars. This may be exacerbated in LSB galaxies, which observations suggest tend to be relatively poor in molecular hydrogen \citep[e.g.,][]{deblok1998co,oneil2003}. If this observed deficiency were linked to lower gas-phase metallicities (see Section \ref{met}) and therefore lower rates of cooling, self-shielding, and dust-grain formation of H$_2$ \citep[e.g.,][]{gerritsen1999,Mihos1999}, the resolution of our simulations would prevent us from seeing its effect on star formation rates.

It is only at the very highest masses (M$_\star\geq$10$^{10}$ M$_\odot$), where bulge+disk LSB galaxies tend to be among the galaxies with the lowest star formation rates, that there is any discernible difference between LSB and HSB galaxy star formation rates in \textsc{Romulus25}. This initially appears counter to observational evidence, which has long suggested that classical LSB galaxies are universally characterized by low star formation rates \citep[e.g.,][]{impey1997}. However, our finding that LSB and late-type HSB galaxies of the same mass have similar star formation rates is in agreement with data from larger samples from e.g., \citet{galaz2011}, \citet{du2015}, \citet{Junais2023}, and \citet{Phillipps2023}, as well as previously simulated samples from EAGLE \citep{Kulier2020} and TNG100 \citep{PerezMontano2022}. It is not that these galaxies have low star formation rates, but rather that they have low star formation efficiencies given their large HI supply \citep[e.g.,][]{boissier2008}.

In the right-hand panel of Figure \ref{fig:SFRlsb}, we show the ratio of each galaxy's HI mass to its star formation rate, using the same samples as in the left-hand panel. This is the ``HI depletion timescale" (t$_\mathrm{HI}$)---that is, the amount of time that it would take each galaxy to consume its entire supply of HI assuming that star formation continued at the current rate indefinitely. In general, we see that a galaxy's t$_\mathrm{HI}$ is inversely correlated with its stellar mass, with our high mass (M$_\star\gtrsim$10$^{10}$ M$_\odot$) galaxies typically having t$_\mathrm{HI}\leq$1 Gyr and the majority of our low mass galaxies (M$_\star<$10$^9$ M$_\odot$) having t$_\mathrm{HI}\gtrsim$ the age of the Universe. Both this trend and the range of t$_\mathrm{HI}$ values spanned by our simulated LSB galaxy samples appear to be consistent with the observed samples. It is also clear that LSB galaxies tend to have considerably longer HI depletion times than HSB galaxies do, particularly for M$_\star\geq$10$^9$ M$_\odot$. The median t$_\mathrm{HI}$ for our LSB galaxies (excluding those galaxies with SFR = 0 M$_\odot$/yr) is 8.6 Gyr, which is $\sim$4.5 times longer than the median t$_\mathrm{HI}$ for our HSB galaxies. As the instantaneous star formation rates of our LSB galaxies are very similar to those of our HSB galaxies, we can conclude that this difference is primarily driven by the differences in the populations' HI reservoirs.

\begin{figure*}
\centering
\includegraphics[width=0.97\textwidth]{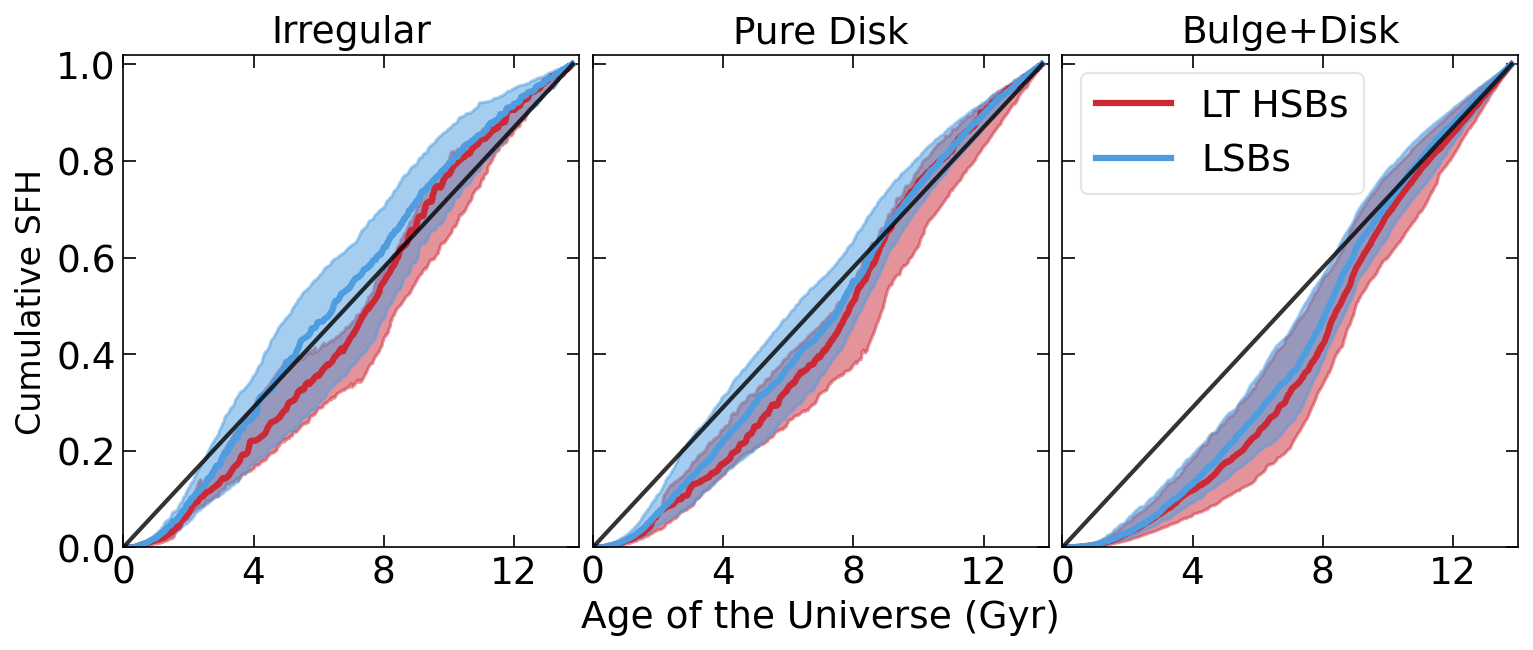}
\caption{Cumulative star formation histories for our simulated classical LSB galaxies (in blue) and mass-matched samples of late-type HSB galaxies (in red). The thick solid lines track the evolution of the median while the shading indicates the interquartile range at each time. The black line indicates the evolutionary track that we would expect to see for a galaxy that formed stars at a constant rate over its entire lifetime. Bulge+disk and pure disk LSB galaxies have star formation histories that are very similar to those of late-type HSB galaxies of similar mass. While both pure disk and irregular LSB galaxies have star formation histories that have been very close to constant over their lifetimes, irregular LSB galaxies tend to have assembled their stellar populations slightly more quickly than similar HSB galaxies.}
\label{fig:csfh}
\end{figure*}

That star formation is not so different in LSB galaxies and HSB galaxies is also evident when we compare their star formation histories. In Figure \ref{fig:csfh}, we plot the median cumulative star formation histories and accompanying interquartile ranges for our classical LSB galaxies. We split our sample into the three subsamples we have previously defined---irregular LSB galaxies, pure disk LSB galaxies, and bulge+disk LSB galaxies---so as to identify differences in their histories and so that we can compare each to an appropriate sample of HSB galaxies from \textsc{Romulus25}.

Because the average HSB galaxy is more massive than the average LSB galaxy and most galaxy properties (anti-)correlate with stellar mass, an unweighted comparison between the groups would make it difficult to disentangle differences related to surface brightness from those linked to mass. In order to construct a comparison sample, we therefore sort all LSB galaxies and late-type central HSB galaxies into bins of 0.125 dex in log stellar mass. To account for any disparities in stellar mass distribution, we assign each HSB galaxy in a given comparison sample a weight corresponding to the fraction of bulge+disk/pure disk/irregular LSB galaxies that occupy the same stellar mass bin. Each of the three comparison samples contains the same galaxies (i.e., every late-type central HSB galaxy), but weights them differently, such that the binned stellar mass distribution of an LSB galaxy subsample is the same as the binned and weighted stellar mass distribution of the corresponding HSB galaxy comparison sample.\footnote{We note that this method prioritizes mass-matching over morphology-matching. This has the potential to be problematic -- particularly when comparing star formation across samples -- due to the strong connection between bulge mass and quenching \citep[e.g.,][]{Bell2008,Bluck2014,Kim2018,Bluck2022}. However, we have repeated our analysis, comparing (non-)bulge+disk LSB galaxies only to (non-)bulge+disk HSB galaxies, and have confirmed that accounting for this potential difference in morphology does not impact our conclusions.} This method successfully produces HSB galaxy comparison samples with stellar mass distributions that are statistically indistinguishable (a two-sample Kolmogorov-Smirnov (KS) test yields p$>$0.1) from those of LSB galaxies for the bulge+disk and pure disk subsamples. The HSB galaxy comparison sample for the irregular LSB galaxies, however, remains slightly discrepant (p$\sim$10$^{-3}$) due to the tendency of the HSB galaxies in a given 0.125 dex bin to be less massive, on average, than the LSB galaxies in that bin at M$_\star<$10$^{8.5}$ M$_\odot$. While we expect the impact of this disparity to be relatively small, it may contribute to (or compensate for) differences between irregular LSB galaxies and their HSB galaxy comparison sample. We use the weighted, mass-matched, late-type central HSB comparison samples throughout the rest of the paper.

\begin{figure*}
\centering
\includegraphics[width=0.97\textwidth]{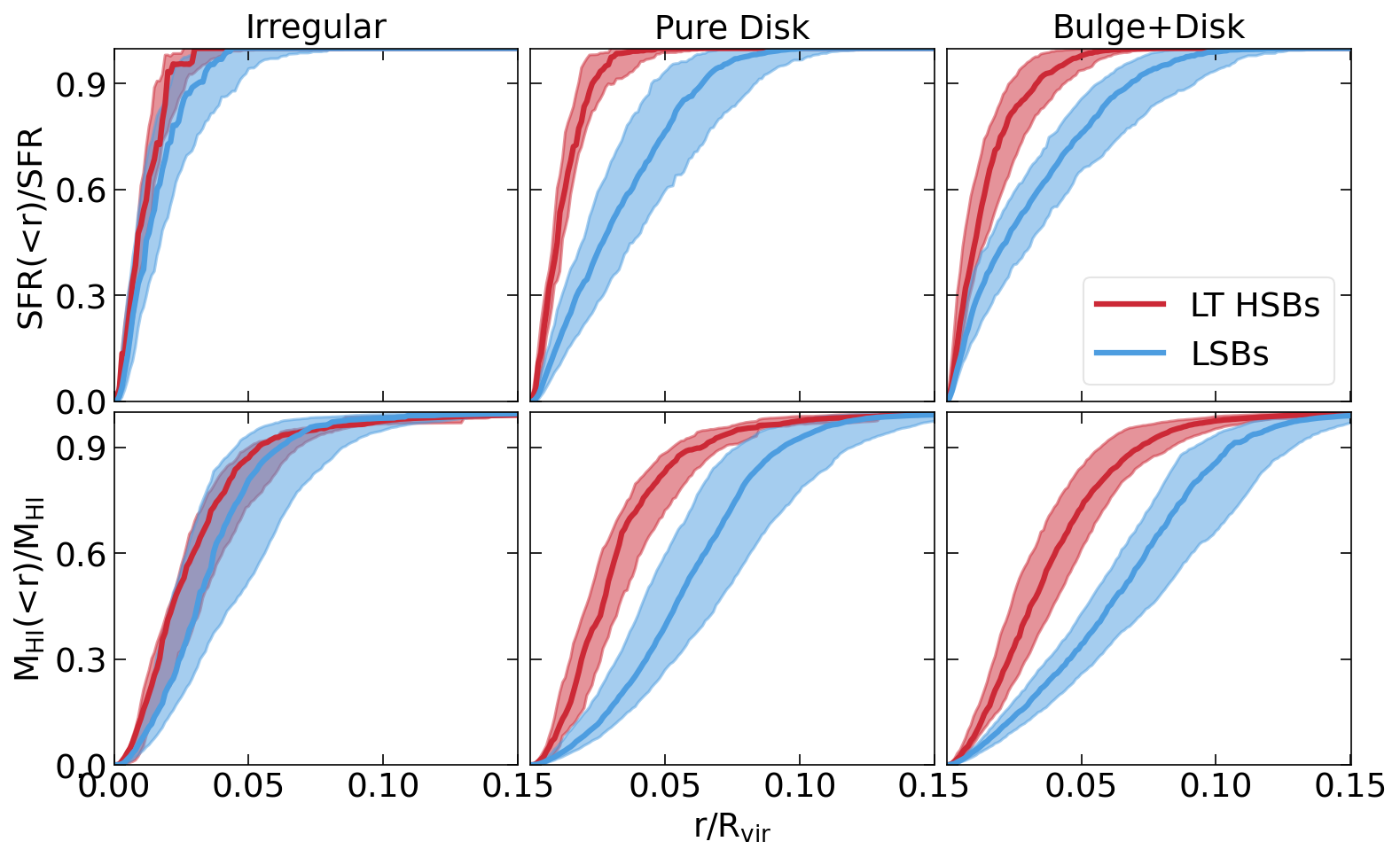}
\caption{Enclosed star formation rate (\textit{top}) and  HI mass (\textit{bottom}) profiles, normalized to individual galaxy totals and scaled by virial radius, for our classical LSB galaxies (blue) and mass-matched samples of late-type HSB galaxies (red). The thick solid lines track the median profile while the shading indicates the interquartile range at each radius. HI and star formation are both considerably more extended in LSB galaxies than in HSB galaxies. The fraction of HI contained within the inner $\sim$5\% of the virial radius is particularly low in pure disk and bulge+disk LSB galaxies.}
\label{fig:profs}
\end{figure*}

We find that there is no discernible difference between the star formation histories of bulge+disk LSB galaxies and their comparison sample and only a very slight tendency for pure disk LSB galaxies to assemble their stellar populations earlier than their HSB counterparts. Both pure disk and irregular LSB galaxies have star formation histories that deviate very little from a constant rate of star formation. This finding is supported by multiple observational studies: it is generally agreed upon that LSB galaxies have formed stars at a low but steady rate for approximately a Hubble time \citep[e.g.,][]{bell2000,wyder2009,schombert2013,schombert2014spitzer,lei2018}. Because of this near-constant star formation rate, we see the greatest difference between the star formation histories of LSB and HSB galaxies in the irregular galaxies, where HSB galaxies tend to grow more slowly until z$\sim$1 and then rapidly catch up to the LSB galaxies.

For pure disk and bulge+disk LSB galaxies, where we see a difference is less in the rate or time at which stars form, but in \textit{where} they form. In the top panels of Figure \ref{fig:profs}, we show cumulative star formation as a function of radius for our classical LSB galaxies and their mass-matched late-type central HSB galaxy counterparts. As in Figure \ref{fig:csfh}, we plot the median profile and the interquartile range for each sample. Profiles are normalized to the total star formation rate of each individual galaxy and scaled by each galaxy's virial radius in order to allow us to directly compare the shapes and extents of profiles across different samples.

In each of our subsamples, star formation is considerably more extended in LSB galaxies than in HSB galaxies. This provides a natural explanation for their low surface brightnesses: if an HSB galaxy and an LSB galaxy have the same stellar mass, the same number of stars will be spread over a larger area in the LSB galaxy, resulting in lower stellar surface densities and therefore lower surface brightnesses. This is very similar to what we see in the field ultra-diffuse galaxies in \textsc{Romulus25} \citep{wright2021}. 

In the bulge+disk sample, we see a slight deficit of star formation in LSB galaxies even within the bulge region. This is also true in an absolute sense: the specific star formation rates within the inner 500 pc of LSB galaxies are lower than those within the inner 500 pc of HSB galaxies. This is not necessarily what we would expect, as bulge-dominated LSB galaxies are traditionally thought to have typical HSB bulges embedded in extended low surface brightness disks that have evolved largely independently of one another \citep[e.g.,][]{Beijersbergen1999,Barth2007,Lelli2010,Morelli2012}. This slight difference between the bulges of HSB galaxies and the bulges of LSB galaxies provides further evidence that there is a greater evolutionary connection between the bulges and disks of individual galaxies, in agreement with observational findings from e.g., \citet{Galaz2006} and \citet{Saburova2021}. We discuss this further in Section \ref{col}.

As we might expect, given that stars form in cold dense gas, the HI profiles of the LSB galaxies in \textsc{Romulus25} are also more extended than those of their HSB counterparts. In the bottom panels of Figure \ref{fig:profs}, we show enclosed HI mass as a function of radius for each of our subsamples. The greater extent of HI disks in LSB galaxies also explains why we might expect to see more sporadic star formation in LSB galaxies, as many have suggested \citep[e.g.,][]{vandenhoek2000,boissier2003,boissier2008,schombert2014spitzer,schombert2014sps}. More of the mass of HI that exists within LSB galaxies is at larger radii, where it is necessarily spread over a larger area. This leads to lower surface densities and correspondingly lower star formation efficiencies. We also see a difference in the shapes of enclosed HI profiles. Particularly in pure disk and bulge+disk LSB galaxies, there is a distinct deficit in central HI relative to their HSB counterparts.

\begin{figure*}
\centering
\includegraphics[width=0.97\textwidth]{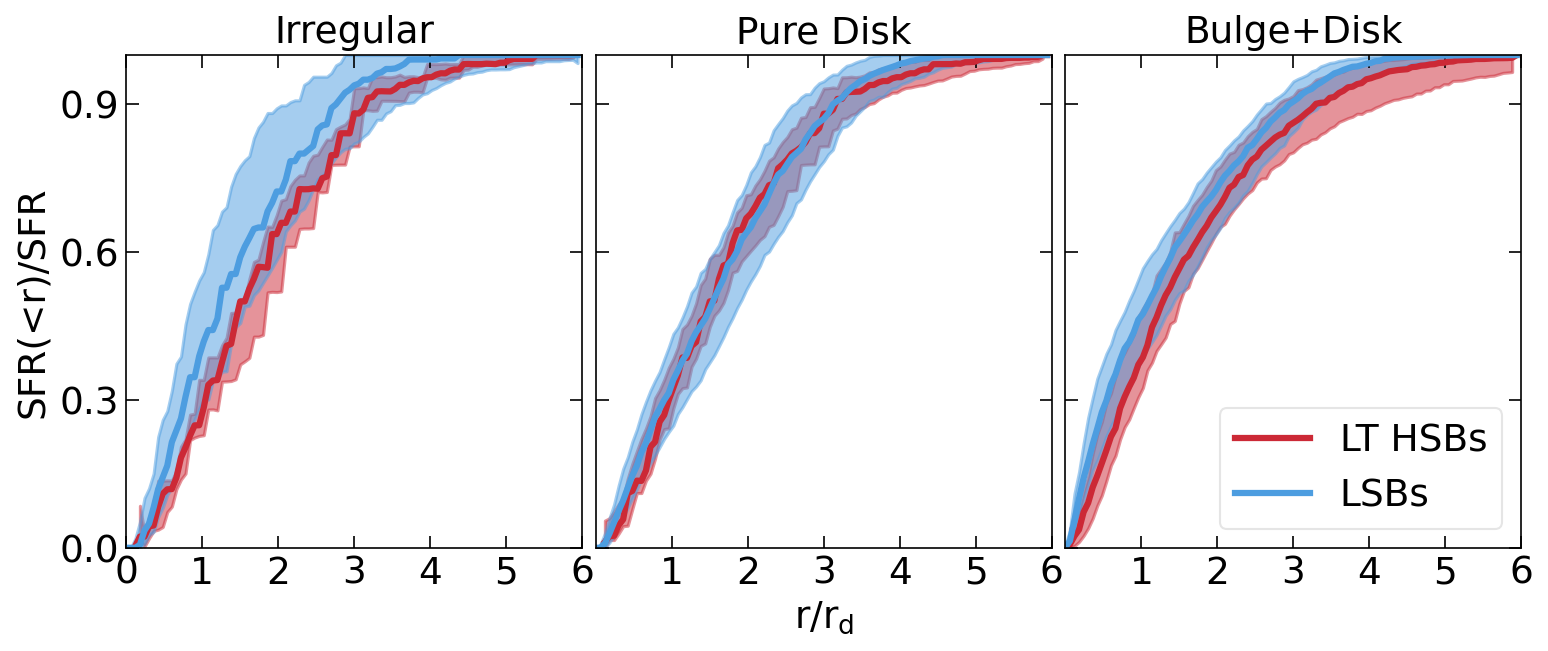}
\caption{Enclosed star formation rate profiles, normalized to individual galaxy totals and scaled by disk scale length ($r_\mathrm{d}$), for our classical LSB galaxies (blue) and mass-matched samples of late-type HSB galaxies (red). The thick solid lines track the median profile while the shading indicates the interquartile range at each radius. Most of the differences that we see between the profiles of LSB and HSB galaxies in Figure \ref{fig:profs} are eliminated by scaling by $r_\mathrm{d}$, rather than by R$_\mathrm{vir}$.}
\label{fig:scprofs}
\end{figure*}

Many of the differences that we see between the profiles of LSB and HSB galaxies are due to the fact that LSB galaxies tend to have larger scale lengths ($r_\mathrm{d}$ in Eq. \ref{eq:diskfit}) than HSB galaxies \citep[e.g.,][]{sprayberry1995TFR,deblok1996HI,du2019}, as we showed in Figure \ref{fig:struct}. If we scale the profiles by $r_\mathrm{d}$, rather than by R$_\mathrm{vir}$, the offset between the star formation rate profiles of LSB and HSB galaxies is greatly reduced. We show this in Figure \ref{fig:scprofs}. The use of $r_\mathrm{d}$ makes the star formation rate profiles of bulge+disk and pure disk LSB galaxies and their HSB counterparts nearly indistinguishable. This makes sense, as the distribution of light, from which we measure the scale length, typically follows the distribution of star formation. 

\subsubsection{Metallicity}
\label{met}
\begin{figure*}
\centering
\includegraphics[width=0.97\textwidth]{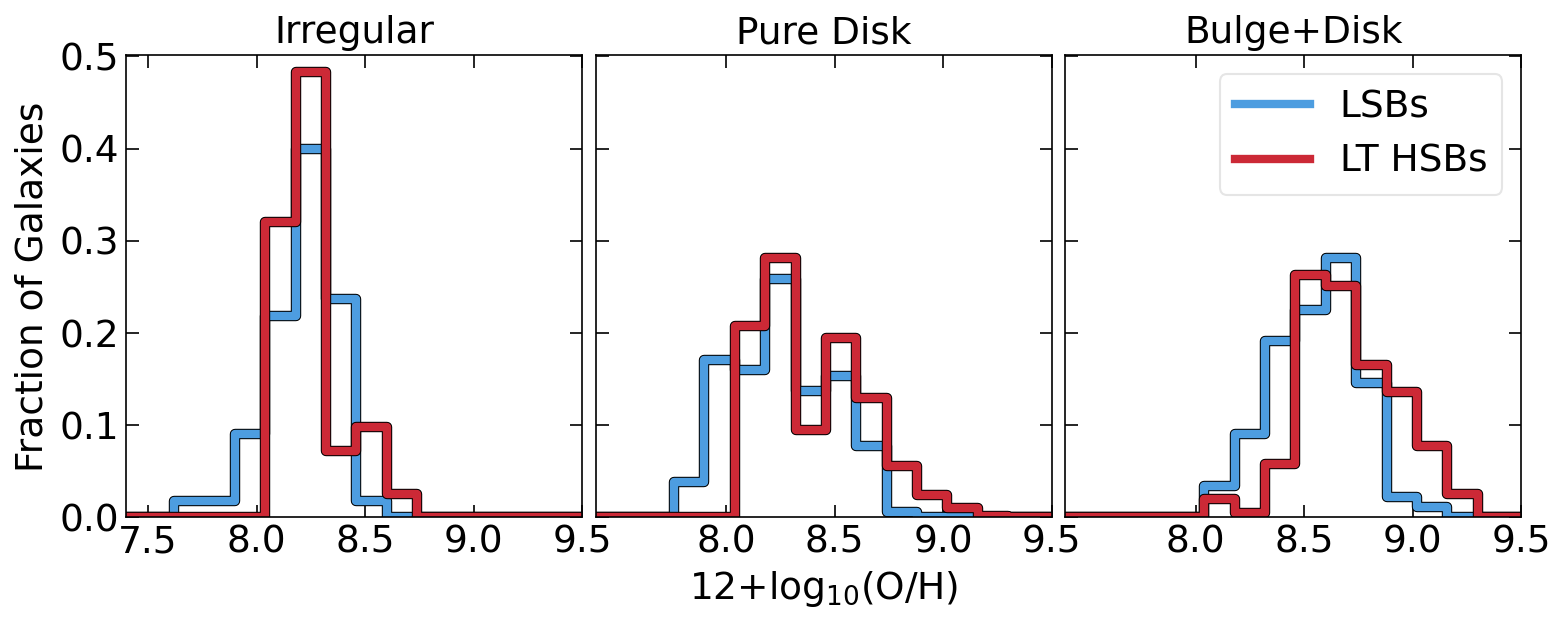}
\caption{Distribution of metallicity, as measured by the oxygen abundance in cold gas, for classical LSB galaxies (shown in blue) and mass-matched samples of late-type HSB galaxies (shown in red). In each of our subsamples, the LSB galaxies tend to have lower metallicities than the HSB galaxies, although this offset is most noticeable in the pure disk and bulge+disk samples.}
\label{fig:methist}
\end{figure*}

Alongside star formation history, metallicity serves as one of the most important probes of the evolution of a galaxy. In Figure \ref{fig:methist}, we show the distributions of metallicities, as measured from the abundance of oxygen in cool gas (T$<$10$^4$ K), for our classical LSB and mass-matched late-type HSB galaxy samples. In all of our subsamples, the HSB galaxies have higher median metallicities than the LSB galaxies. The gap is largest in the bulge+disk sample, where the difference in median metallicity between late-type HSB and LSB galaxies is 0.12 dex, but persists even in the irregular sample, where the difference in median metallicity is 0.02 dex. In all instances, a two-sample KS test rejects the hypothesis that the samples of late-type LSB and HSB galaxies are drawn from the same distribution with p $<$ 0.02. These findings are in line with observations, which have long measured lower gas-phase metallicities in LSB galaxies than in HSB galaxies of the same stellar mass \citep[e.g.,][]{mcgaugh1994oxygen,kuziodenaray2004,Young2015,Cao2023}, as well as other simulations in which LSB galaxies have been found to have lower stellar metallicities \citep[e.g.,][]{Kulier2020,Tang2024}.

\begin{figure}
\centering
\includegraphics[width=0.47\textwidth]{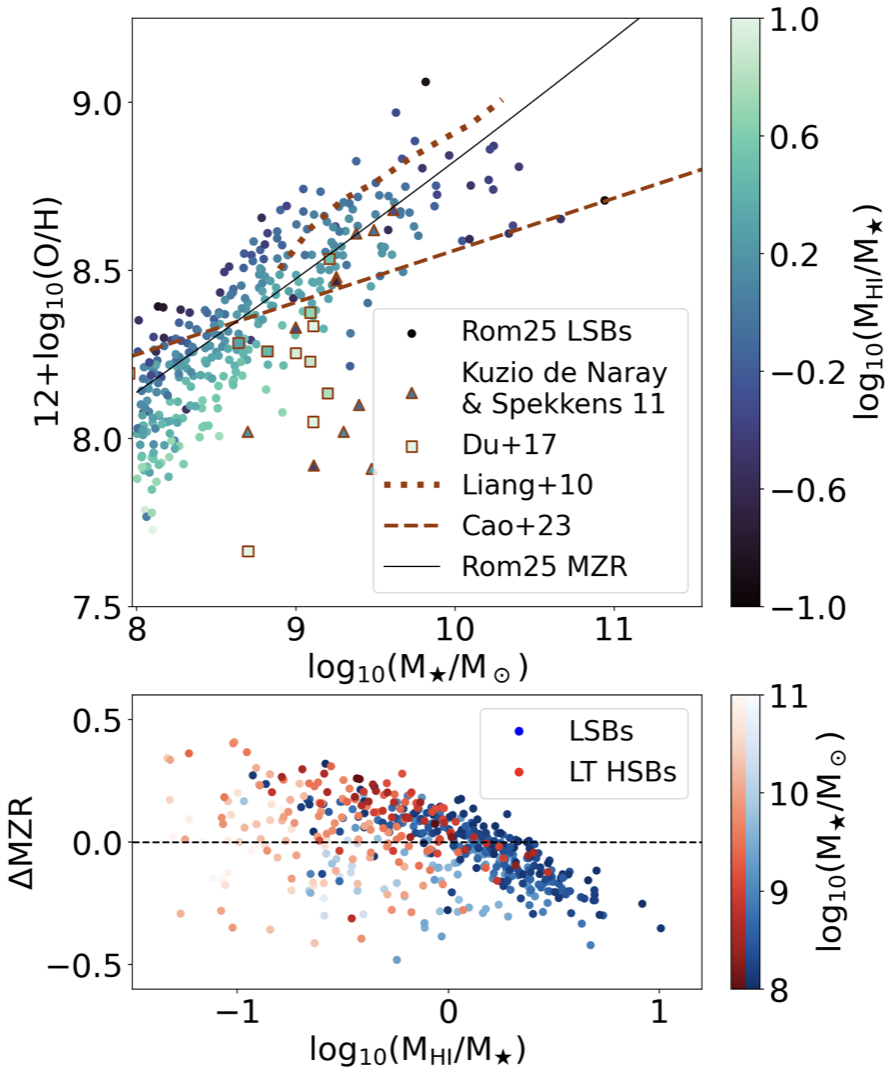}
\caption{\textit{Top:} Mass-metallicity relationship for LSB galaxies in \textsc{Romulus25}, compared to data from observations of LSB galaxies from \citet{kuziodenaray2011}, \citet{Du2017}, \citet{Liang2010}, and \citet{Cao2023}. Points for individual galaxies are colored by HI-richness (M$_\mathrm{HI}$/M$_\star$). \textit{Bottom:} HI-richness of both LSB galaxies and late-type HSB galaxies in \textsc{Romulus25} plotted against each galaxy's distance from the mass-metallicity relationship for late-type central \textsc{Romulus25} galaxies and colored by stellar mass. In both LSB and HSB populations, more HI-rich galaxies tend to be more metal-poor, although this relation is tighter for lower mass galaxies.}
\label{fig:massmet}
\end{figure}

In the top panel of Figure \ref{fig:massmet}, we compare the stellar masses and gas-phase metallicities of our classical LSB galaxies to those of several observed LSB galaxy samples. While the \textsc{Romulus25} (shown as circles), \citet[][dotted line]{Liang2010}, and \citet[][dashed line]{Cao2023} galaxies clearly follow a mass-metallicity relation (MZR), the small samples from \citet[][triangles]{kuziodenaray2011} and \citet[][squares]{Du2017} display no such trend and, in fact, lie well below the other three samples. We can likely attribute the bulk of these discrepancies to the extreme difficulty of properly calibrating metallicities derived via different indicators \citep{Kewley2008}. \citet{kuziodenaray2011} average over the (primarily strong line-based) metallicities calculated from spectra of multiple H~{\sc ii} regions in each LSB galaxy, while \citet{Du2017} and \citet{Cao2023} both adopt samples of edge-on LSB galaxies with metallicities derived from the N2 diagnostic. \citet{Liang2010}'s sample consists of face-on LSB galaxies from SDSS and uses abundances from \citet{tremonti2004}, which are calculated from a combination of many emission lines. We note that we have shown only the median relation from their ``vLSB'' galaxy sample here, as the surface brightness criterion used to define it (i.e., $\mu_{0,B}\geq$22.75) most closely resembles our own LSB galaxy definition. While all of these studies conclude that LSB galaxies are metal-poor in comparison to HSB galaxies, the only one that includes a uniform analysis of both LSB and HSB galaxies is \citet{Liang2010}, which finds only a small difference in the median MZRs of LSB and late-type HSB galaxies, with the disparity being most evident among galaxies with 9.5$\leq$log$_\mathrm{10}$(M$_\star$/M$_\odot$)$\leq$10.5. Both this qualitative finding and the more quantitative MZR shown in Figure \ref{fig:massmet} are consistent with what we see in \textsc{Romulus25}, although we caution against overinterpreting the latter given the calibration challenges referenced above.

The relatively low metallicities observed in classical LSB galaxies were originally interpreted as evidence that they were unevolved and had perhaps initiated star formation later than HSB galaxies \citep[e.g.,][]{mcgaugh1994oxygen}. However, subsequent observations revealed the presence of substantial ancient stellar populations in LSB galaxies \citep[e.g.,][]{Galaz2002,Bergmann2003,schombert2013,schombert2014sps,schombert2014spitzer}, ruling out a significantly atypical formation epoch. A more popular explanation for the metal deficiency of LSB galaxies stems from their gas-richness, reflecting the widely observed tendency for star-forming galaxies across a broad range of masses and environments to have gas-phase metallicities that anti-correlate with their abundance of cold gas \citep[e.g.,][]{Hughes2013,Bothwell2013,Zahid2014,Scholte2024}. This relation is generally understood to be a product of low star formation efficiencies maintaining a large supply of under-enriched gas---particularly at the low mass end \citep{brooks2007}---and/or recently accreted metal-poor gas diluting the metal-enriched ISM and thereby simultaneously increasing the gas-richness of a galaxy and decreasing its metallicity \citep[e.g.,][]{Finlator2008,Brisbin2012,Salim2014,Young2015}. As the process of accretion typically increases the gas surface density of the galaxy and triggers a starburst and thus metal-rich outflows, the MZR--M$_\mathrm{HI}$/M$_\star$ relation and the more well-known anti-correlation between metallicity and specific star formation rate \citep[the ``fundamental metallicity relation''; e.g.,][]{Ellison2008,Mannucci2010} are often interpreted as different facets of the same underlying relation \citep[e.g.,][]{Bothwell2013}.

We demonstrate that gas-richness is anti-correlated with gas-phase metallicity in our simulated LSB galaxies by shading the points in the top panel of Figure \ref{fig:massmet} by M$_\mathrm{HI}$/M$_\star$. At any given stellar mass, those galaxies that are most gas-rich tend to also have the lowest metallicities. We show this more explicitly in the bottom panel of Figure \ref{fig:massmet}, where we plot M$_\mathrm{HI}$/M$_\star$ for each galaxy against that galaxy's distance from the median MZR ($\Delta$MZR) for central late-type galaxies in \textsc{Romulus25}. Nearly every galaxy with M$_\mathrm{HI}$/M$_\star>$1 lies below the MZR, while the vast majority of galaxies with M$_\mathrm{HI}$/M$_\star<$1 reside above it. There is also a clear anti-correlation between log$_\mathrm{10}$(M$_\mathrm{HI}$/M$_\star$) and $\Delta$MZR that is followed by both LSB and HSB galaxies, which are shown in shades of blue and red, respectively, according to their stellar masses. While the HSB galaxies are more scattered around this relation than the LSB galaxies, this appears to be primarily driven by the fact that the late-type HSB galaxy population includes a larger fraction of high-mass galaxies, which do not conform as well to the relation. Our results therefore indicate that LSB galaxies in \textsc{Romulus25} are relatively metal-poor for the same reason that they are gas-rich.

\begin{figure}
\centering
\includegraphics[width=0.47\textwidth]{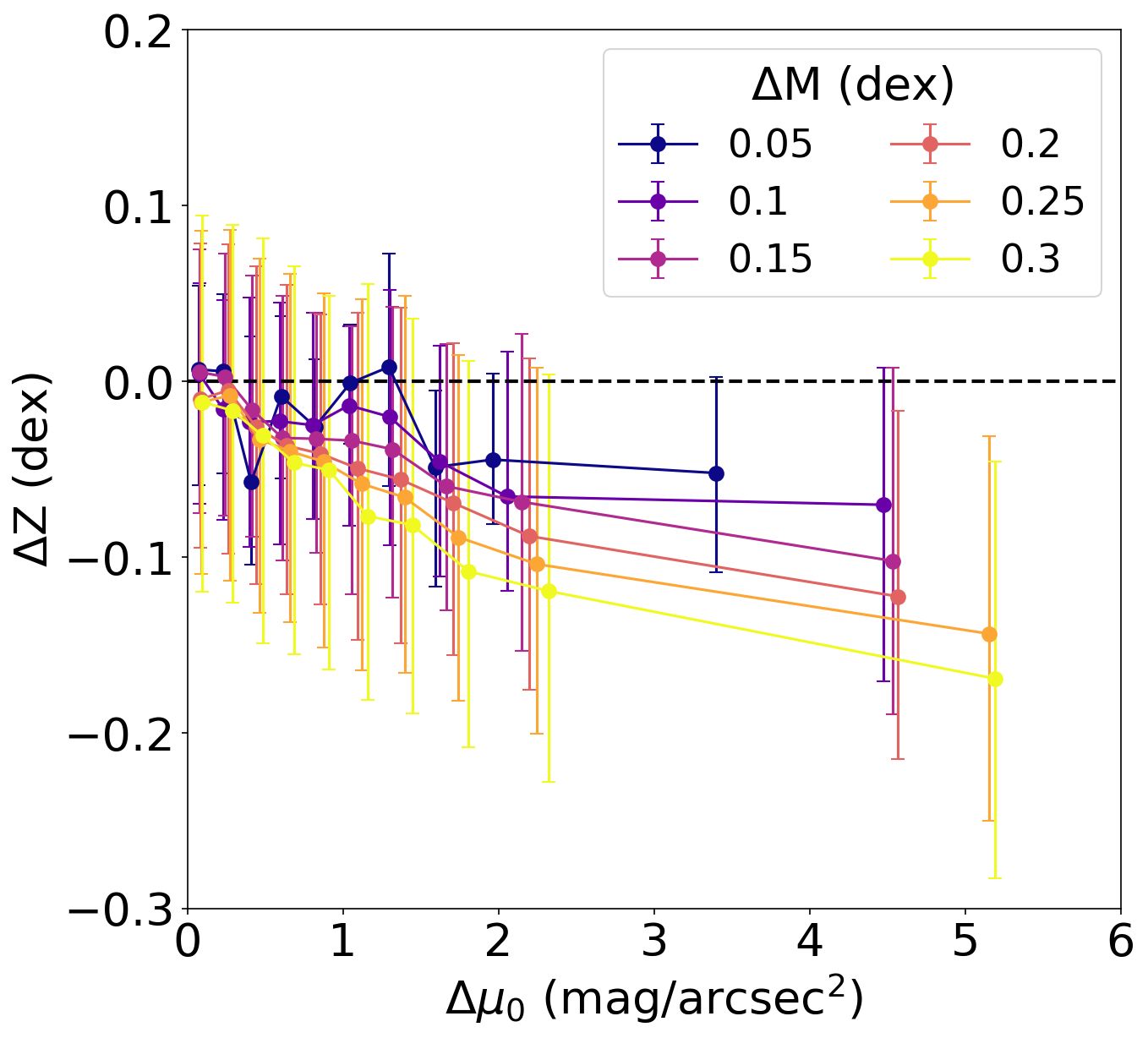}
\caption{The differences in central surface brightness ($\Delta\mu_\mathrm{0}$) and gas-phase metallicity ($\Delta$Z) between any two late-type central galaxies in \textsc{Romulus25} that are less than $\Delta$M dex apart in both M$_\star$ and M$_\mathrm{HI}$. $\Delta$Z and $\Delta\mu_\mathrm{0}$ are calculated by subtracting the values for the brighter galaxy in the pair from those of the fainter galaxy. Data for each value of $\Delta$M are placed in 10 equally populated bins of $\Delta\mu_\mathrm{0}$, with a circle marking the median $\Delta$Z for the bin and error bars indicating the interquartile range. The fainter one galaxy is relative to another, the more comparably metal-poor it tends to be. However, this trend is mostly driven by differences in M$_\star$ and M$_\mathrm{HI}$ and flattens when we reduce the allowed mass differences. Compare to Figure 3 of \citet{Wu2015}.}
\label{fig:sbvmet}
\end{figure}

\citet{Wu2015}, however, argue that lower surface brightness is correlated with lower metallicity even when both stellar mass and gas mass have been taken into account. Using a small sample of observed LSB and HSB galaxies, they showed that, in pairs of galaxies with less than 0.3 dex difference in both M$_\star$ and M$_\mathrm{gas}$, the fainter one galaxy was relative to another, the more comparably metal-poor it tended to be. We repeat this analysis using our sample of central late-type LSB and HSB galaxies from \textsc{Romulus25} and plot the results in Figure \ref{fig:sbvmet}. Following their procedure, we identify pairs of galaxies with similar HI and stellar masses and calculate their relative difference in both central surface brightness ($\Delta\mu_\mathrm{0}$; plotted on the x-axis) and gas-phase metallicity ($\Delta$Z; plotted on the y-axis), subtracting the values for the brighter galaxy from those of the fainter galaxy. For the sake of clarity, we plot only summary values, sorting the galaxy pairs into 10 equally populated bins of $\Delta\mu_\mathrm{0}$ and plotting the median (circle) and interquartile range (error bars) of $\Delta$Z for the galaxy pairs in that bin \citep[cf. Figure 3 of][]{Wu2015}. If we consider pairs that vary by $\leq$0.3 dex in both M$_\star$ and M$_\mathrm{HI}$, as \citet{Wu2015} did, we recover their trend: a larger difference in central surface brightness is correlated with a larger deficit in metals in the fainter galaxy. However, our larger sample allows us to place better constraints on this relationship by reducing the amount by which pairs of galaxies are permitted to differ in mass. We plot summary data for different values of $\Delta$M (the maximum stellar or HI mass by which pairs of galaxies are allowed to differ) in different colors in Figure \ref{fig:sbvmet}. As the mass window narrows, the trend flattens. This suggests that the apparent relationship is primarily driven by mass differences (and therefore gas-richness) within the permitted mass range. For $\Delta$M $\leq$ 0.3 dex and $\Delta\mu_\mathrm{0}\geq$ 1, the lower surface brightness galaxy in a pair will have lower stellar mass and/or be more gas-rich 80\% of the time (compared to 60\% of the time for $\Delta$M $\leq$ 0.05 dex). This yields a median difference in log$_\mathrm{10}$(M$_\mathrm{HI}$/M$_\star$) of 0.1--0.2 dex, which Figure \ref{fig:massmet} indicates corresponds to a similar difference in gas-phase metallicity. While there is still some hint of a correlation between lower surface brightness and lower metallicity in even our narrowest mass window, it is consistent with a slope of 0. Gas-richness is clearly the primary driver of metal deficiency in LSB galaxies in \textsc{Romulus25}. 

\subsubsection{Color}
\label{col}
Another regime in which LSB and HSB galaxies are purported to differ is color \citep[e.g.,][]{pildis1997,schombert2014spitzer}. The combination of blue optical colors, which are typically indicative of a dominant young stellar population, and the low-to-normal star formation rates actually observed in LSB galaxies is a long-standing puzzle with a number of proposed solutions ranging from extremely metal-poor stellar populations to very specific star formation histories. \citet{schombert2014spitzer} show that the range of IR and optical colors in their sample of observed LSB galaxies can be reproduced by stellar population models in which the star formation rate has increased or decreased by a factor of 4 over the last 500 Myr. This finding aligns with previous theories that LSB galaxies experience episodic star formation that, while constant when averaged over large timescales, follows the periodic accretion of gas from the intergalactic medium \citep[e.g.,][]{schombert1990,Galaz2002}---a process that we've already noted has also been invoked to explain their low metallicities. \citet{gerritsen1999} and \citet{oneil2000} note that this scenario implies the existence of a population of red LSB galaxies very similar to the blue ``classical'' LSB galaxies with the exception of being currently entirely quiescent.

\citet{mcgaugh1996} observed early on that the LSB galaxy surveys of the 1990's were likely biased against the discovery of red LSB galaxies due to the use of blue-sensitive plates---a bias that would be perpetuated through the continued use of LSB catalogs compiled with these plates \citep[e.g.,][]{schombert2011,schombert2014spitzer,mcgaugh2017}. Indeed, red LSB galaxies have since been discovered in large numbers, first by \citet{oneil1997}, but increasingly frequently over the last several years by surveys specifically designed to probe the LSB sky \citep[e.g.,][]{greco2018illuminating,Tanoglidis2021,Greene2022,Thuruthipilly2024}. These galaxies were initially discovered almost exclusively in dense environments, but the environments of recently discovered red LSB galaxies are uncertain and there is evidence to suggest that some are genuinely isolated \citep[e.g.,][]{Prole2021}. It is therefore unclear whether classical LSB galaxies are truly unusually blue or if this is merely a selection effect.

In agreement with classical observations, our LSB galaxies are quite blue: $>$ 95\% have B-V $\leq$ 0.65 and the LSB sample as a whole has a median B-V $=$ 0.49. These values are very similar to those found in large samples of observed LSB galaxies from \citet{mcgaugh1994structure}, \citet{burkholder2001}, and \citet{du2015}, all of which have average B-V $=$ 0.49-0.52. That being said, this relative blueness is largely by design: in selecting our late-type samples (both LSB and HSB galaxies), we have excluded all galaxies with B-V $>$ 0.72.

\begin{figure*}
\centering
\includegraphics[width=0.97\textwidth]{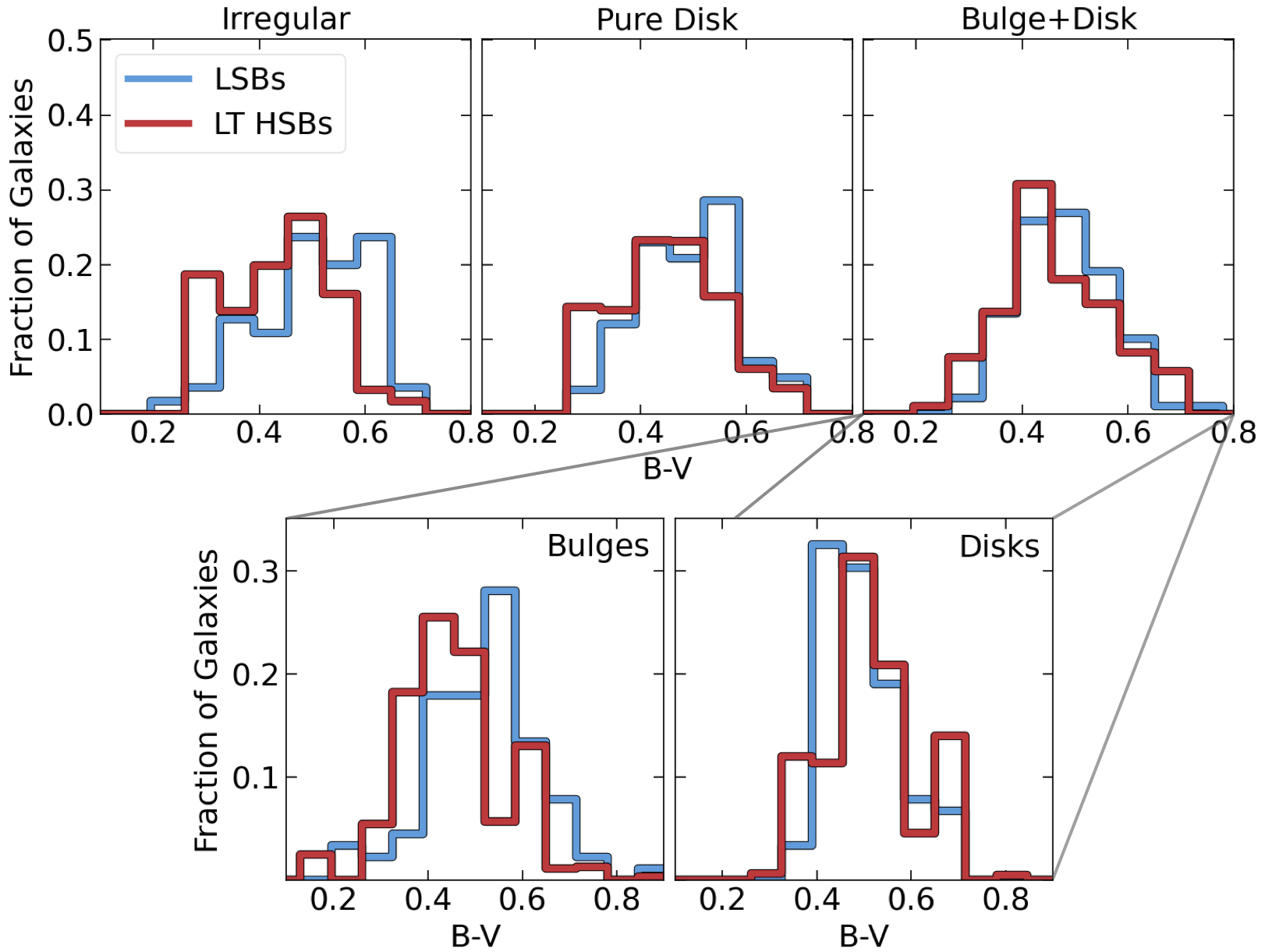}
\caption{\textit{Top:} Distribution of B-V colors for classical LSB galaxies (shown in blue) and mass-matched samples of central late-type HSB galaxies (shown in red). Although our simulated LSB galaxies are relatively blue, the irregular and pure disk samples are, on average, redder than late-type HSB galaxies of similar mass. \textit{Bottom:} Distribution of B-V colors for the bulges and disks of the bulge+disk LSB (blue) and late-type HSB (red) samples. The bulges of the LSB galaxies are redder than those of the HSB galaxies on average.}
\label{fig:bvhist}
\end{figure*}

We show the distribution of B-V colors for our LSB galaxy samples and mass-matched samples of HSB galaxies in Figure \ref{fig:bvhist}. In all of our subsamples, the median B-V color of the LSB galaxies is actually redder than that of the HSB galaxies by 0.03-0.07, with the overall distributions being statistically distinct (p $<$ 0.01 for a two-sample KS test) in all but the bulge+disk sample. For this subsample, we must separate each galaxy out into its bulge and disk components (see Section \ref{struct}), which we do in the bottom panels of Figure \ref{fig:bvhist}, in order to observe a difference between LSB and HSB galaxies. The bulges of late-type LSB galaxies (median B-V $=$ 0.52) are considerably redder than those of late-type HSB galaxies (median B-V $=$ 0.43), while the disks are slightly bluer (median $\Delta$B-V $=$ 0.02). However, only the former is statistically significant (p $<$ 10$^{-4}$). 

The relative colors of our simulated LSB galaxies are in line with what we would expect given their star formation histories (see Figure \ref{fig:csfh}): LSB galaxies in the irregular and pure disk samples tend to build up their stellar populations slightly earlier than HSB galaxies of similar mass, leading to (on average) older and therefore optically redder stellar populations. While there is no discernible difference between the global star formation histories of LSB and HSB galaxies in the bulge+disk sample, it is clear from Figure \ref{fig:profs} that the central HI content and star formation rates of LSB galaxies in this sample are lower than those of HSB galaxies. As we know that the overall bulge masses of these samples are the same (see Figure \ref{fig:struct}), we can infer that LSB galaxies built up the stellar mass that makes up their bulges earlier than HSB galaxies, producing their redder optical colors. Observational evidence of divergent bulge properties in these populations is mixed. Early observational work suggested little difference between the colors of LSB and HSB galaxy bulges \citep[e.g.,][]{Beijersbergen1999,Galaz2006}, but more recent resolved studies have found that H~{\sc ii} regions tend to avoid the central regions of LSB galaxies and some extreme examples appear to form stars exclusively at the edges of their disks \citep[e.g.,][]{schombert2013,Young2015,Young2020UGC628,Young2020UGC8839}.

More generally, those galaxies in \textsc{Romulus25} with instantaneous star formation rates that lie below (above) their lifetime average (M$_\star$/t$_\mathrm{Hubble}$) do tend to be redder (bluer). However, this is equally true for LSB and HSB galaxies, which we have already shown have similar star formation rates for their stellar masses. This is largely unsurprising, as B-V color is known to primarily be a reflection of recent star formation \citep[e.g.,][]{Schombert2001}.

However, another potentially significant factor that we have not yet considered is the effect of metallicity on the colors of our galaxies. While the isochrones \citep{Marigo2008,Girardi2010} that we use to calculate the colors of individual star particles are selected from an age-metallicity grid, this is fairly simplistic. \citet{schombert2014sps} showed that correcting for the impact of metallicity on the IMF and the relative frequency of e.g., blue horizontal branch stars and blue stragglers, could make model LSB galaxy stellar populations bluer by $\Delta$B-V$\approx$0.1. Because of the uncertainties associated with these stellar population models, we have not attempted to incorporate these effects into our calculations. However, the larger impact from metallicity is likely to be in the form of internal reddening due to dust. Dust is thought to condense out of metals and the mass of dust relative to gas in a given galaxy is therefore often modeled as a strong function of metallicity \citep[e.g.,][]{Remy-Ruyer2014}. Accordingly, we might expect LSB galaxies, which we know to be metal-poor for their stellar masses, to also be dust-poor. Observations appear to support this idea: direct detections of dust in LSB galaxies are rare \citep[e.g.,][]{wyder2009,Junais2023} and extinction is generally found to be extremely low \citep[e.g.,][]{mcgaugh1994oxygen,deblok1998oxygen}. Consequently, the effects of dust are often entirely neglected in analyses of LSB galaxy stellar populations and their colors \citep[e.g.,][]{schombert2014sps}. 

\begin{figure*}
\centering
\includegraphics[width=0.97\textwidth]{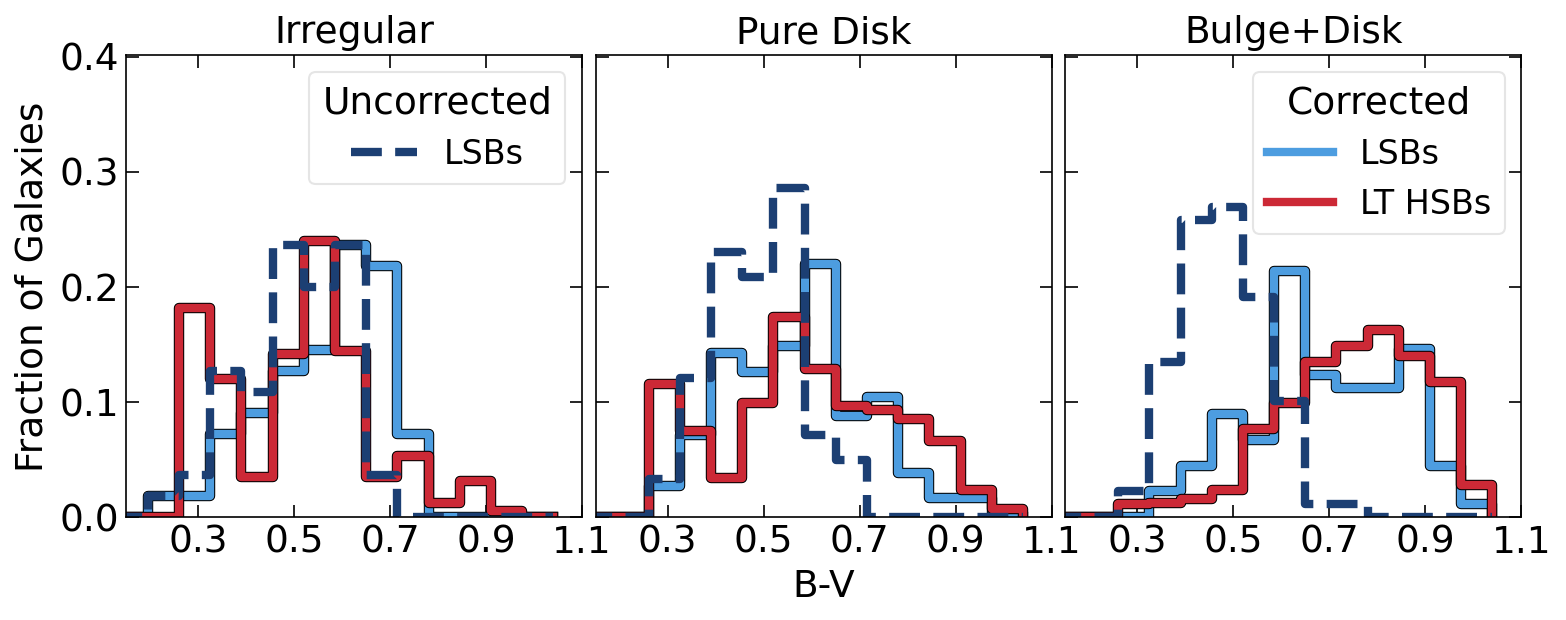}
\caption{The distributions of B-V colors for classical LSB galaxies (shown in blue) and mass-matched samples of central late-type HSB galaxies (shown in red). Distributions shown with solid lines are corrected for internal reddening due to dust following \citet{Shivaei2020}. Because dust is often entirely neglected in LSB galaxy modeling, we also include a dashed blue line showing the uncorrected B-V colors of these galaxies. The higher gas-phase metallicities of HSB galaxies cause them to appear redder.}
\label{fig:bvdust}
\end{figure*}

In Figure \ref{fig:bvdust}, we reproduce the top panels of Figure \ref{fig:bvhist}, but correct the B-V color of each galaxy for the effects of dust. We adopt the best fit gas-phase metallicity--stellar reddening relation from Figure 9 of \citet{Shivaei2020}, which is based on data from the MOSDEF survey and covers a similar range in metallicity (log(O/H)+12=8--8.8) to our own sample. Overall, the HSB galaxies experience considerably more reddening than the LSB galaxies. In the pure disk and bulge+disk groups, the effect is strong enough to make the median LSB galaxy bluer than the median HSB galaxy by 0.01 and 0.1, respectively, with the overall B-V distributions being statistically distinct (p $<$ 0.01 for a two-sample KS test). While the irregular LSB galaxies are still slightly redder than their HSB counterparts, the two distributions are no longer distinguishable. If we instead assume, as many do, that the colors of LSB galaxies are entirely unaffected by dust due to e.g., the low surface densities of both their stars and their gas \citep[e.g.,][]{McGaugh1997}, even the irregular LSB galaxies become bluer than their mass-matched HSB comparison sample. For reference, we include the distribution of uncorrected B-V colors of the LSB galaxies as a dashed line in Figure \ref{fig:bvdust}. The reality likely lies somewhere between the corrected and uncorrected distributions. Our simulations suggest that, if LSB galaxies are, indeed, bluer than HSB galaxies even once selection effects have been taken into account, it is because their lower metallicities result in less internal reddening due to dust, in agreement with multiple previous observational studies \citep[e.g.,][]{vandenhoek2000,wyder2009,schombert2011}.

\subsection{Formation of Classical LSB Galaxies}
\label{form}

\subsubsection{Environments}
\label{env}
Some of the earliest proposals regarding the formation of LSB galaxies centered on their environments \citep[e.g.,][]{lacey1991,hoffman1992,zaritsky1993,bothun1993}. If LSB galaxies are less clustered than HSB galaxies, it is likely that they would have had fewer encounters with other galaxies, leading to fewer tidally triggered bursts of star formation and allowing them to maintain extended gaseous disks \citep[e.g.,][]{herpich2014}. Evidence for this is mixed. A number of authors have found that LSB galaxies tend to have fewer neighbors than HSB galaxies on relatively small scales: $\leq$0.5 Mpc \citep{zaritsky1993}, $\leq$1 Mpc \citep{galaz2011,honey2018}, and $\leq$2-3 Mpc \citep{bothun1993,mo1994}. However, reduced clustering on larger scales is not well established. While a number of authors \cite[see, for example,][]{bothun1986,thuan1991,schombert1992} have failed to find any evidence that LSB galaxies are isolated at scales $\geq$ 2 Mpc, \cite{rosenbaum2009} claim that their investigation of over 2000 LSB galaxies from SDSS suggests that LSB galaxies have (on average) fewer neighbors than HSB galaxies at scales as large as 5 Mpc. More recently, \citet{du2015} found no evidence that the environments occupied by LSB galaxies in the $\alpha$.40-SDSS DR7 survey were any different than those inhabited by other HI-rich galaxies.

\begin{figure*}
\centering
\includegraphics[width=0.97\textwidth]{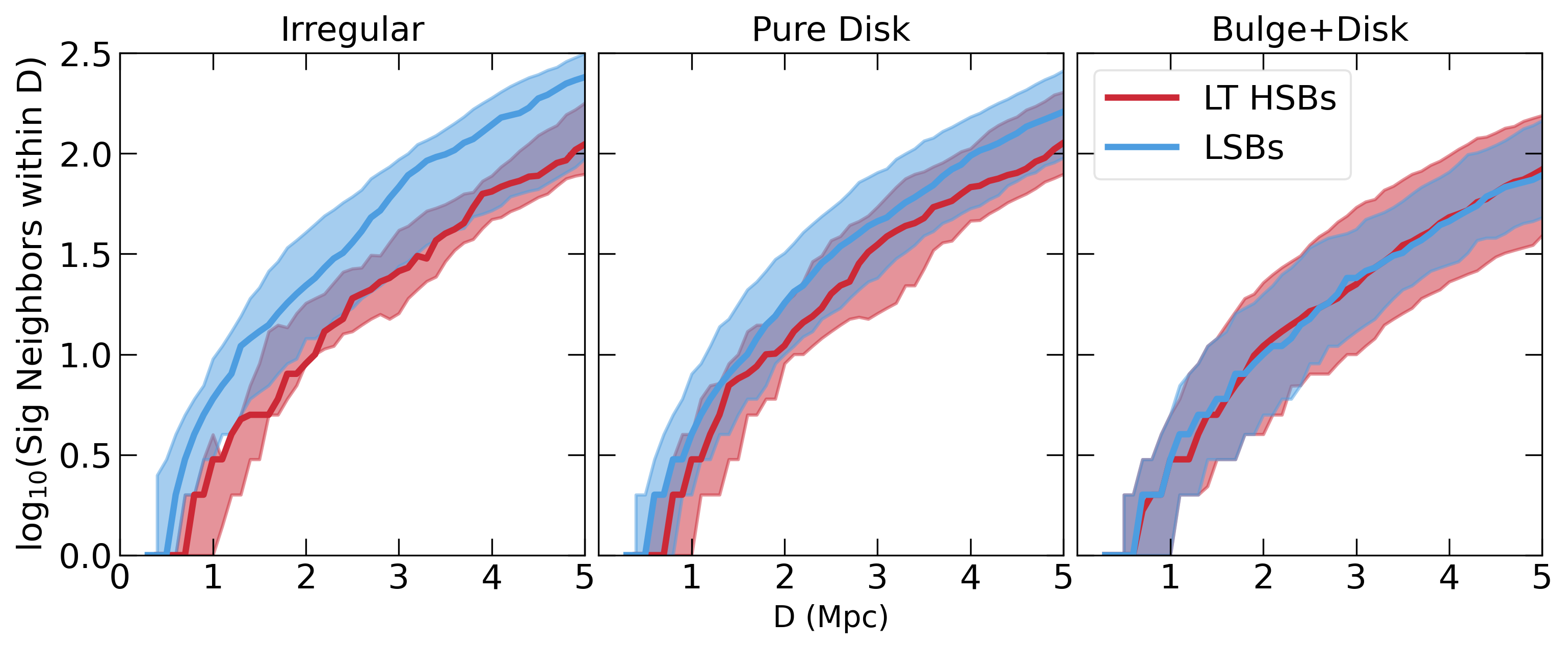}
\caption{The number of significant neighbors within a distance D of our classical LSB galaxies (shown in blue) and late-type HSB galaxies (shown in red), where a significant neighbor is any resolved galaxy with M$_\mathrm{vir}$ at least 10\% that of the LSB or HSB galaxy we are searching around. Thick lines indicate median values at each distance and shaded regions show the interquartile range. While bulge+disk LSB galaxies inhabit the same environments as HSB galaxies, pure disk LSB galaxies are slightly more likely to be found in denser environments on scales $\gtrsim$1.5 Mpc and irregular LSB galaxies on average have more neighbors than their HSB counterparts at all scales, albeit with a high amount of scatter.} 
\label{fig:env}
\end{figure*}

In Figure \ref{fig:env}, we show the number of significant neighbors within distances of up to 5 Mpc around our LSB galaxies and their mass-matched  late-type HSB galaxy comparison samples. Here, a significant neighbor is defined to be any central galaxy that meets our resolution criterion (i.e., M$_\mathrm{vir} >$ 3$\times$10$^9$ M$_\odot$) and has M$_\mathrm{vir}$ at least 10\% that of the LSB or HSB galaxy we are searching around. We do not find any evidence that bulge+disk LSB galaxies inhabit different environments than HSB galaxies of similar stellar mass. However, there is a slight tendency for pure disk LSB galaxies to have more significant neighbors than their HSB counterparts at distances $\gtrsim$1.5 Mpc and irregular LSB galaxies inhabit denser environments, on average, than HSB galaxies at all of the scales we consider, albeit with significant scatter. 

This being said, we must reiterate that \textsc{Romulus25} is a relatively small volume and contains neither galaxy clusters nor true voids. We do not sample the full range of environments observed in the real universe and our results on this front---particularly at larger scales, where we necessarily have a fairly limited sample---should be viewed cautiously. However, our finding that bulge+disk LSB galaxies (and, to some extent, pure disk LSB galaxies) do not inhabit abnormal environments for their stellar mass is consistent with observational results from \citet{Blanton2005} and \citet{du2015}, as well as simulations from \citet{Kulier2020} and \citet{Perez-Montano2024}, who find that central LSB galaxies in the \textsc{EAGLE} and \textsc{IllustrisTNG} simulations typically inhabit the same environments as central HSB galaxies of similar mass. Additionally, there is some indication that low density environments may only be necessary for the formation and maintenance of giant LSB galaxies \citep[e.g.,][]{hoffman1992}, which, as we have noted, we do not expect to form in a volume the size of \textsc{Romulus25}.

The fact that we tend to find irregular LSB galaxies in slightly higher density environments than HSB galaxies of the same stellar mass at $z=0$ suggests that these galaxies may form through the process described in \citet{martin2019}. Using the \textsc{Horizon-AGN} simulation, \citet{martin2019} found that cumulative perturbation by fly-bys and the ambient tidal field is the primary mechanism driving the divergent evolution of LSB and HSB galaxies. In a follow-up paper, \citet{Jackson2021} found that low-mass LSB galaxies in \textsc{NewHorizon}, a zoom-in simulation of a region of \textsc{Horizon-AGN}, are also the products of environment, forming in regions of higher dark matter density where they are able to accrete gas more rapidly and therefore experience more star formation and associated supernova feedback at early epochs. This process, combined with increased tidal perturbation at lower redshifts, results in lower surface brightness, more extended galaxies. In order to compare to their results, we follow them in quantifying the tidal perturbation that a given galaxy experiences using the perturbation index (PI):
\begin{equation}
PI = \int_{z=3}^{z=0} \sum_i \frac{M_i}{M_\mathrm{gal}} \Bigl( \frac{r_\mathrm{gal}}{D_i}\Bigl)^3 \,dt\,
\end{equation}
where $M_\mathrm{gal}$ and $r_\mathrm{gal}$ are the mass and radius of the LSB or HSB galaxy in question, and we are summing over the masses of ($M_\mathrm{i}$) and distances to ($D_\mathrm{i}$) all galaxies (including satellites) within a 3 cMpc radius sphere of our galaxy at each timestep from $z=3$ to $z=0$. However, we alter their method slightly, using M$_\mathrm{200}$, rather than M$_\star$, for $M_\mathrm{i}$ and $M_\mathrm{gal}$, in order to account for the drop-off in baryonic content in lower mass dark matter halos. Additionally, we adopt 0.1R$_\mathrm{200}$, rather than $r_\mathrm{eff}$ (or $r_\mathrm{d}$), for $r_\mathrm{gal}$ in order to isolate an increase in PI due to tidal perturbations from an unrelated change in disk size. While the disk of a galaxy with larger $r_\mathrm{eff}$ will experience more perturbation than a more compact disk in the same environment, the direction of causation --- i.e., whether increased perturbation is producing a larger, lower surface brightness disk, or whether a larger, lower surface brightness disk is experiencing more perturbation purely due to its size --- is difficult to establish if we set PI$\propto$r$_\mathrm{eff}^3$. In particular, because $r_\mathrm{eff}$ and $r_\mathrm{d}$ are systematically larger in LSB galaxies than in HSB galaxies, using them would cause the same underlying perturbation to yield a higher PI for an LSB than an HSB galaxy. We therefore use 0.1R$_\mathrm{200}$ to avoid any systematic bias in our determination of whether tidal perturbations play a role in LSB galaxy formation.

\begin{figure*}
\centering
\includegraphics[width=0.97\textwidth]{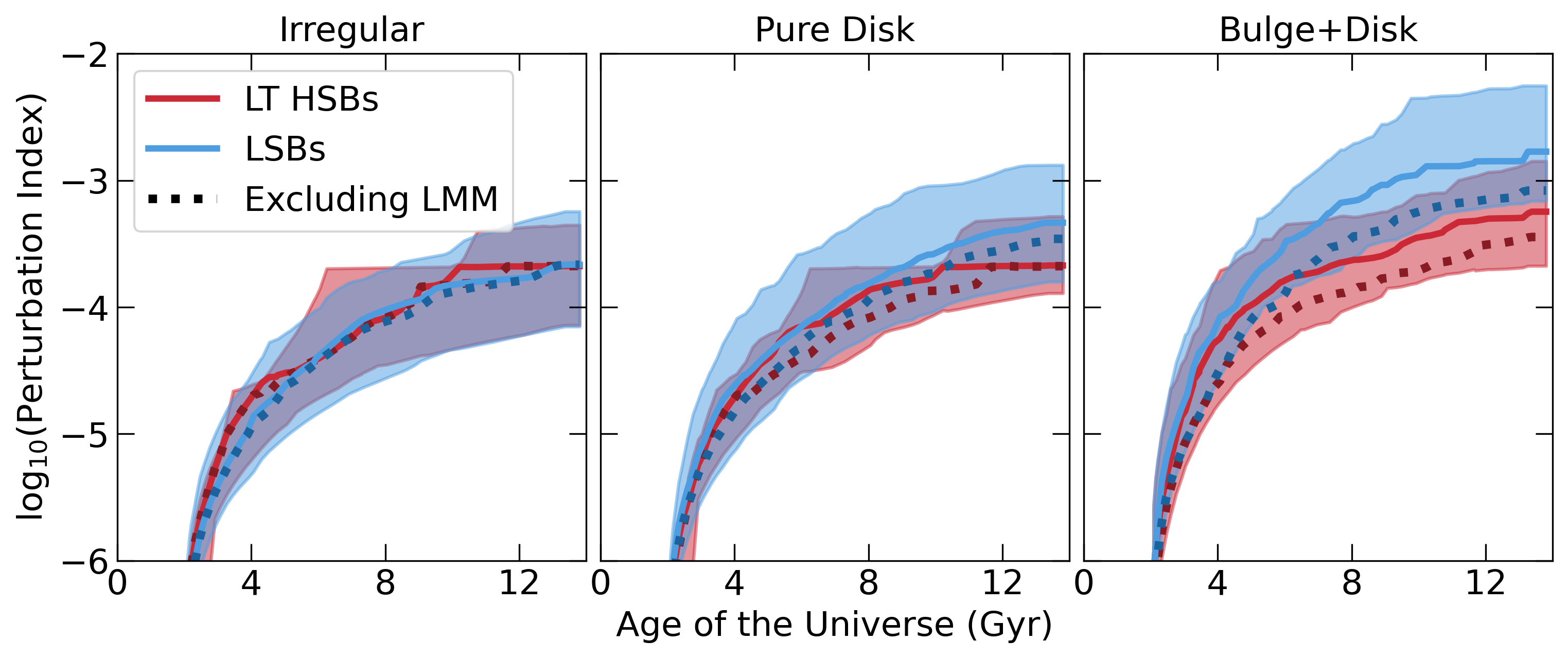}
\caption{The evolution of the perturbation index of LSB galaxies (shown in blue) and late-type HSB galaxies (shown in red). The thick solid lines track the median evolution while the shading indicates the interquartile range at each time. In dark blue and dark red dotted lines, we also show the median values for the perturbation index when the contribution from the galaxy's last major merger has been excluded. While there is a tendency for pure disk and disk+bulge LSB galaxies to experience more tidal perturbation than HSB galaxies, this is largely driven by each galaxy's last major merger.} 
\label{fig:PI}
\end{figure*}

We show the evolution of PI for each of our LSB and HSB galaxy subsamples in Figure \ref{fig:PI}. Note that, because PI is an integrated quantity, we include only those galaxies that are reliably traced back to $z=3$ (90\% of the sample) in this plot. By this metric, irregular LSB galaxies do not experience more perturbation than HSB galaxies. Rather, the ambient tidal fields of the two populations are quite similar on average. Bulge+disk LSB galaxies and, to a lesser extent, pure disk LSB galaxies, have higher PIs than their HSB counterparts, particularly at $z<1$ ($t\gtrsim$6 Gyr). However, this difference is primarily driven by a single tidal interaction, particularly for pure disk LSB galaxies. As shown in dotted darker lines in Figure \ref{fig:PI}, excluding the contribution from each galaxy's final major merger brings the LSB and HSB galaxies into much closer agreement for reasons that we will explore in the next section. There is still a small signature of excess tidal perturbation in bulge+disk LSB galaxies even when this final merger is neglected. We will investigate the role of unbound interactions in LSB galaxy evolution in future work (Corillo in prep.).

Perhaps it is ultimately unsurprising that the LSB galaxies in \textsc{Romulus25} do not predominantly form via the same processes as those in \textsc{Horizon-AGN} and \textsc{NewHorizon}. The LSB galaxies (both classical LSB galaxies and ultra-diffuse galaxies) studied by \citet{martin2019} are primarily distinguished from the HSB galaxies in their sample by their relative dearth of star-forming gas and correspondingly low star formation rates, even when only field galaxies are considered. By contrast, our LSB galaxies are HI-bearing (by definition) and form stars at similar rates to the HSB galaxies we compare them to. Additionally, while our irregular and predominantly low-mass LSB galaxies are in slightly denser environments at $z=0$, they appear to have only recently fallen into them. Their mass accretion histories are indistinguishable from those of late-type HSB galaxies and their PIs are actually lower until the last $\sim$1 Gyr of the simulation. We are therefore likely studying a different population of LSB galaxies than \citet{martin2019} and \citet{Jackson2021}.  

\subsubsection{Angular Momentum and Mergers}
\label{mergers}
The global merger histories of the classical LSB galaxies and late-type HSB galaxies in \textsc{Romulus25} are very similar. They have similar numbers of mergers---both major and minor---and the distributions of the mass ratios and gas contents of their most significant mergers are indistinguishable. The only notable difference is that bulge+disk LSB galaxies tend to have experienced their last major merger slightly more recently (a median of 9.52 Gyr ago) than late-type HSB galaxies of the same stellar mass (a median of 10.53 Gyr ago). Note that throughout this analysis, we define a major merger to be one in which the merging galaxy has M$_\mathrm{200}$ at least 20\% that of the primary galaxy, following \citet{wright2021}.

\begin{figure*}
\centering
\includegraphics[width=0.84\textwidth]{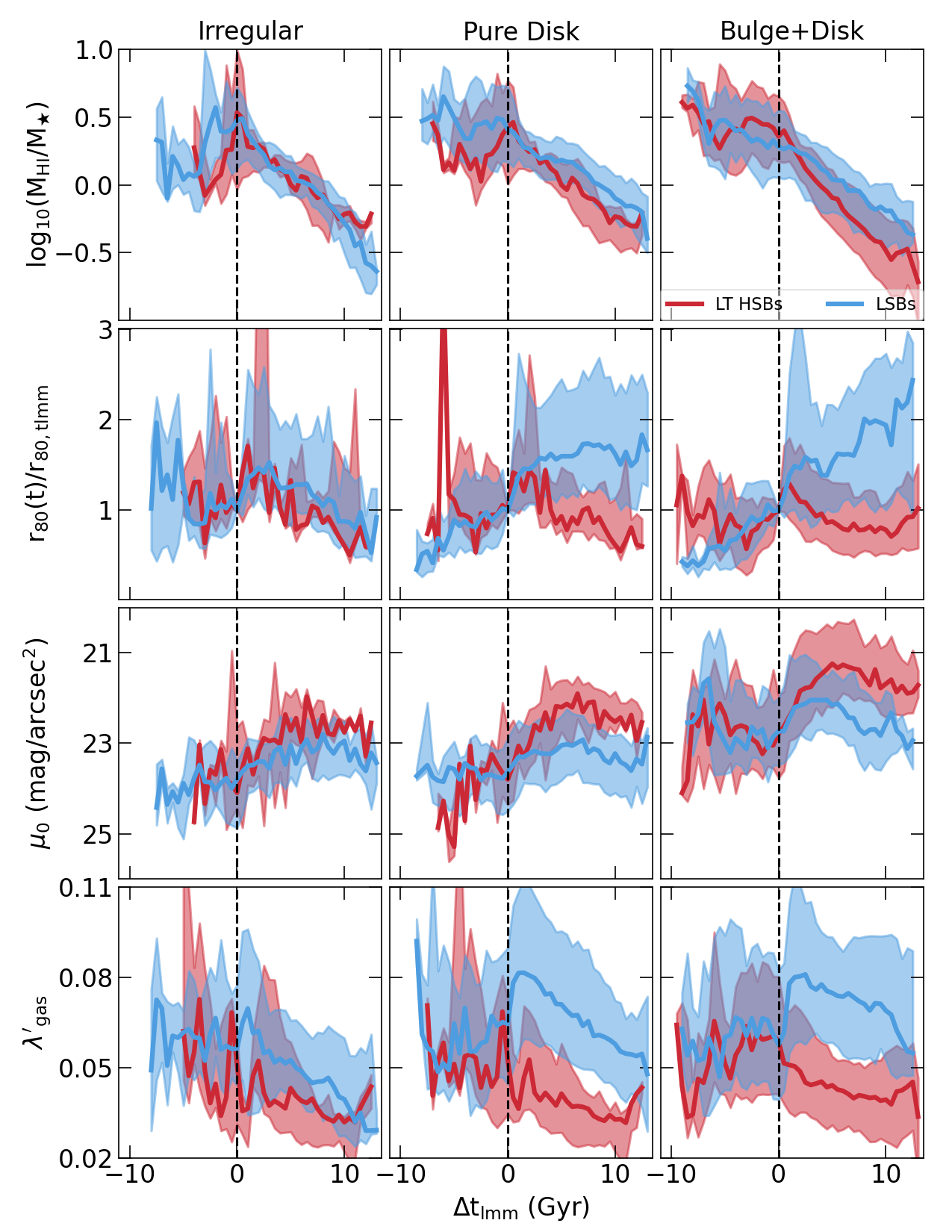}
\caption{HI-to-stellar-mass fraction, radius within which 80\% of star formation is contained (scaled by its value at the time of the last major merger), B-band central surface brightness, and gas spin as a function of time relative to that of the last major merger for LSB galaxies and late-type HSB galaxies in our three groups. Negative time values indicate the time until the merger takes place, while positive values indicate the time that has passed since it occurred. The dashed black vertical lines at $\Delta$t$_{\mathrm{lmm}}=0$ Gyr show the time at which the merger begins. The thick colored lines denote the median evolutionary track for each subsample, while the shading indicates the interquartile range. Mergers spin up LSB galaxies and cause star formation to move outward relative to its position in HSB galaxies, resulting in fainter centers. In HSB galaxies, mergers often cause spin-down and result in more rapid depletion of star-forming gas.}
\label{fig:evorelmer2}
\end{figure*}

Despite the similarities in their global merger histories, LSB and HSB galaxy progenitors react very differently to major mergers. We examine the impacts of major mergers on the evolution of LSB and late-type HSB galaxies in Figure \ref{fig:evorelmer2}. In the top three panels, we have plotted evolutionary tracks for the traits that we have found most distinguish our LSB galaxies from their HSB counterparts: gas-richness (measured here by M$_\mathrm{HI}$/M$_\star$)---which we've shown is also the driver of their relative metal deficiency, spatially extended star formation (measured here by r$_{80}$, the radius within which 80\% of the galaxy's total star formation is occurring), and low central surface brightness (measured here by $\mu_0$, the B-band central surface brightness corresponding to the exponential disk fit). Each evolutionary track is scaled relative to the time at which each galaxy last experienced a major merger ($\Delta$t$_{\mathrm{lmm}}$)---defined as the time at which the virial radii of the primary and secondary galaxies first overlap, following \citet{hetznecker2006}. Negative time values therefore indicate the time until the merger takes place, while positive time values indicate the time that has elapsed since the merger occurred.

Although LSB galaxies and HSB galaxies initially have similar M$_\mathrm{HI}$/M$_\star$, this ratio rapidly decreases in HSB galaxies following their mergers. While we also see a drop-off in gas-richness in LSB galaxies, it isn't as extreme for those in the pure disk and bulge+disk groups. HSB galaxies tend to experience a jump in central specific star formation rate in the time following the mergers, which likely leads to gas outflows, heating, and earlier gas consumption. By contrast, central star formation rates typically decline relatively steadily in LSB galaxies beginning soon after the mergers and continuing until the end of the simulation. This is true in all 3 subsamples, but is least prominent in the irregular LSB galaxy group, which is likely why they are not particularly gas-rich compared to HSB galaxies of the same stellar mass.

Looking at the second row of panels, we can see that mergers are, indeed, altering the locations at which star formation occurs within LSB galaxies. Following the mergers, we see an outward migration of star formation in LSB galaxies relative to HSB galaxies. This is most pronounced in the bulge+disk and pure disk groups, where we know from Figure \ref{fig:profs} that star formation rate and HI profiles differ most strongly. We have chosen to scale r$_{80}$ by its value at the time of the last major merger in order to highlight the fact that, while HSB galaxies typically experience a brief expansion in r$_{80}$ immediately following their mergers and then return to their original extent (in some cases even becoming more compact), star formation in pure disk and bulge+disk LSB galaxies often continues to expand long after the mergers are over, extending to 2-3 times its original radius even 10 Gyr later.

The third row of panels reveals a direct consequence of this redistribution of star formation away from the centers of LSB galaxies and into their outskirts: fainter central surface brightnesses. While the post-merger jump in central star formation rate causes the centers of HSB galaxies to brighten, the mergers have little immediate effect on the centers of LSB galaxies. Rather, as star formation at the centers of these galaxies continues to dwindle, their central stellar populations become, on average, older and therefore optically fainter. This is also why we see redder bulges in bulge+disk LSB galaxies than we do in late-type HSB galaxies. Although we've chosen to show the central surface brightness of the disk fit here, we find that the evolution of the bulge surface brightness is very similar: star-forming gas and star formation both level off in the bulges of LSB galaxies (while increasing in the bulges of HSB galaxies) following their mergers, resulting in the older, redder bulges we see in Figure \ref{fig:bvhist}.

The explanation for the divergent post-merger evolutionary paths of LSB and HSB galaxy progenitors appears to be related to another factor that has long been thought to play a role in LSB galaxy formation: angular momentum. Large-scale N-body simulations have shown that, during major mergers, the orbital angular momentum of the secondary galaxy is converted to the internal spin of the remnant galaxy, typically increasing the spin of the primary galaxy by 25-30\% \citep[e.g.,][]{vitvitska2002,hetznecker2006}, but there is significant scatter. In the bottom panel of Figure \ref{fig:evorelmer2}, we plot evolutionary tracks for specific angular momentum, parameterized by gas spin \citep[$\lambda_\mathrm{gas}'$;][]{bullock2001}. Note that this differs from the original \citet[][]{peebles1969} $\lambda$ in that it lacks any explicit dependence on energy, and should therefore be independent of redshift, making it more appropriate for the time series analysis we present here. Prior to experiencing their last major mergers, LSB galaxies and HSB galaxies have very similar spin distributions. Following their mergers, however, LSB galaxies experience substantial spin-up, while HSB galaxies are actually substantially spun down. Although these differences are most dramatic immediately after the mergers in most of the samples, the disparity persists through the end of the simulation, leaving the LSB galaxies with elevated spins at $z=0$ (although we do see some spin-down in irregular LSB galaxies -- and spin-up in the HSB comparison sample -- at $t_\mathrm{lmm}>$10 Gyr that may reflect the slightly denser environments that irregular LSB galaxies inhabit at late times). Because more than 80\% of the median galaxy's stars are formed during or after this final major merger, the alteration in gas spin leaves a permanent imprint on the distribution of the galaxy's stars. While the spin-down of HSB galaxies results in star formation becoming more compact, the spin-up of LSB galaxies produces much more spatially extended star formation, which we see in Figure \ref{fig:evorelmer2} as a growth in r$_{80}$.

LSB galaxies have long been theorized to be the natural occupants of high angular momentum dark matter halos \citep[][]{dalcanton1997}. Various semi-analytic models \citep[e.g.,][]{jimenez1998,Boissier2000,Prantzos2000,boissier2003,Cervantes-Sodi2009}, N-body simulations \citep[e.g.,][]{mayer2004,bailin2005,maccio2007}, and idealized simulations \citep[e.g.,][]{Churches2001} have confirmed that the colors, radii, surface brightnesses, and chemical compositions of LSB galaxies can be reproduced in model galaxies with higher-than-average specific angular momentum. A number of modern cosmological simulations have also shown that high spin is a driving characteristic of these galaxies \citep[e.g.,][]{dicintio2019,Kulier2020,PerezMontano2022,Zhu2023,Perez-Montano2024,Stoppacher2025}. Although the spin of dark matter halos cannot be directly measured in observed LSB galaxies, calculations of the specific angular momentum of the gas and stars, as well as estimates of the halo spin based on models of the dark matter mass profile suggest that the angular momenta of both the baryonic and dark components of these galaxies are elevated relative to those of HSB galaxies \citep[e.g.,][]{PerezMontano2019env,ManceraPina2020,Salinas2021}---something that we also see in our sample. Several authors have also identified a correlation between spin parameter and HI content \citep{Cervantes-Sodi2009,Huang2012spin,obreschkow2016,lagos2017,stevens2018,zoldan2018,ManceraPina2021,Hardwick2022,Liu2025}, suggesting a further connection between LSB galaxies and high spin halos.

However, that mergers---rather than large-scale tidal torquing---might be responsible for this high spin is a relatively new theory. \citet{dicintio2019} found that, in a small sample of zoom-in simulations, those galaxies that had experienced major mergers in which the angular momentum vector of the gas of the primary galaxy was more closely aligned with the orbital angular momentum vector of the secondary were more likely to be spun-up and experience a decrease in central surface brightness and an increase in effective radius. They reported that this was the primary method by which classical LSB galaxies were formed in the \textsc{NIHAO} simulations.

\citet{wright2021} and follow-up work in \citet{VanNest2022} showed that the population of isolated ultra-diffuse galaxies (UDGs) in \textsc{Romulus25} formed via major mergers. While they found that the overall interaction histories of UDGs and mass-matched non-UDGs were quite similar, they noted that the mergers that produced UDGs tended to happen earlier and to significantly increase the spin of the primary galaxy. As a result, the star formation within the disk of the galaxy became more spread out and the central surface brightness dropped while the effective radius grew. Although \citet{wright2021} found no substantial differences between the orbital configurations of the mergers that produced UDGs and those that resulted in more compact HSB galaxies, they theorized that this was likely due to the masses of the galaxies, all of which were dwarfs (M$_\star<$10$^9$M$_\odot$). \citet{dicintio2019} proposed that while the evolution of massive galaxies is governed by angular momentum, that of dwarf galaxies is primarily driven by feedback.

\begin{figure*}
\centering
\includegraphics[width=0.97\textwidth]{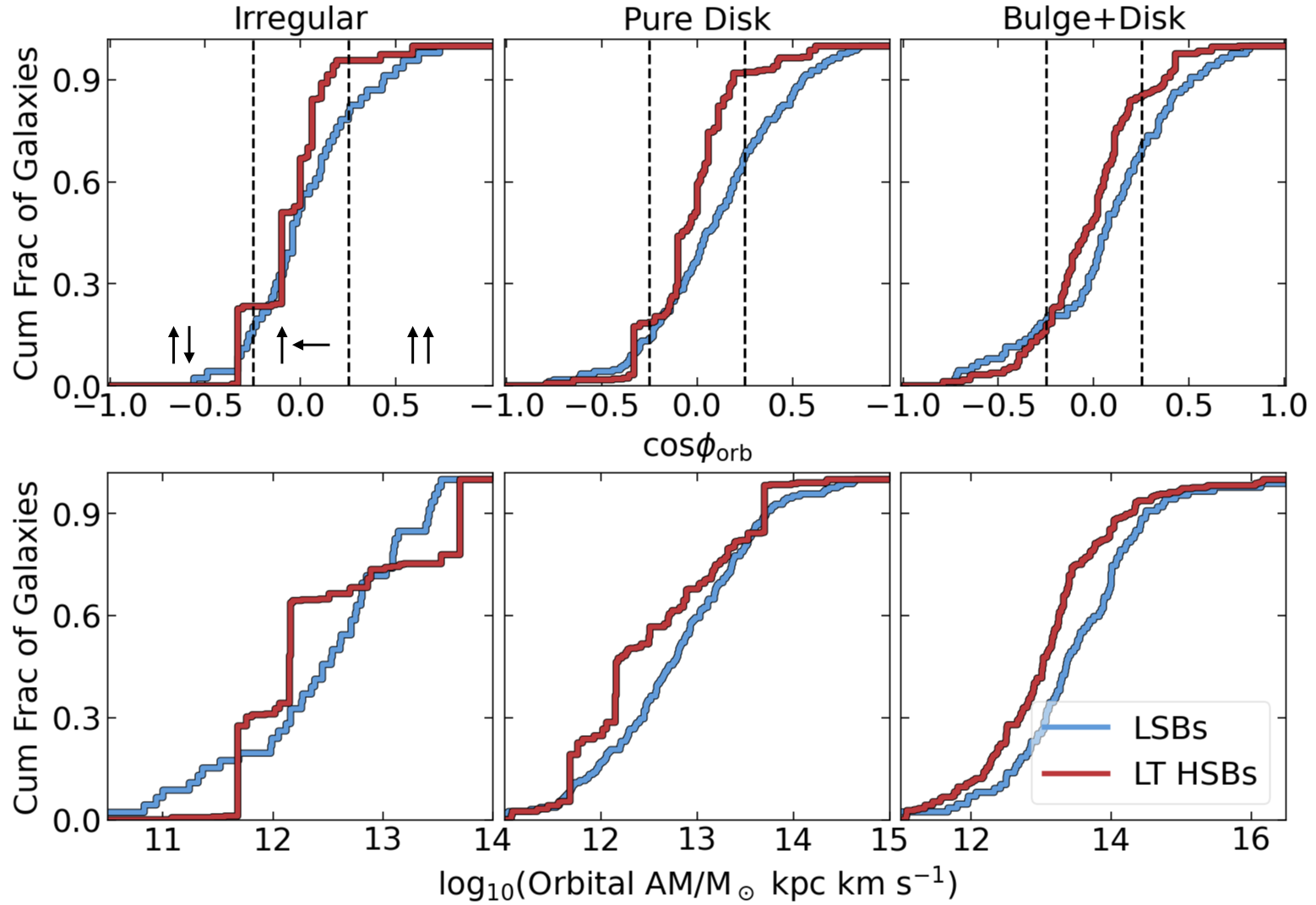}
\caption{Distributions of merger orientations (\textit{top}) and orbital angular momenta (\textit{bottom}) for our classical LSB galaxies and mass-matched samples of late-type HSB galaxies. $\phi_\mathrm{orb}$ is the angle between the orbital angular momentum vector of the secondary galaxy and the specific angular momentum vector of the gas of the primary galaxy for the last major merger. Accordingly, cos$\phi_\mathrm{orb} =$ 1 indicates a perfectly co-rotating, co-planar merger, cos$\phi_\mathrm{orb} =$ -1 indicates a perfectly counter-rotating, co-planar merger, and cos$\phi_\mathrm{orb} =$ 0 indicates a perfectly perpendicular merger. Orbital angular momentum is that of the secondary galaxy at infall. LSB galaxies in all three categories are more likely to have experienced higher orbital angular momentum mergers and/or co-rotating mergers than their HSB counterparts, and less likely to have experienced perpendicular mergers (-0.25$\leq\mathrm{cos}\phi_\mathrm{orb}\leq$0.25, indicated by dashed vertical lines in the top panels).}
\label{fig:lsbmerg}
\end{figure*}

We might, therefore, expect to see a greater dependence of galaxy properties on merger configurations in the predominantly more massive galaxies studied here. In the top panels of Figure \ref{fig:lsbmerg}, we show the distributions of last major merger orientations for our classical LSB galaxies and mass-matched samples of late-type HSB galaxies. Following \citet{dicintio2019}, merger orientation is characterized by $\phi_\mathrm{orb}$---the angle between the orbital angular momentum vector of the secondary galaxy ($\vec{J}_\mathrm{orb} = m\vec{r}\times\vec{v}$) and the specific angular momentum vector of the gas of the primary galaxy ($\vec{J}_\mathrm{gas,primary}$) at infall (i.e., t$_\mathrm{lmm}$):
\begin{equation}
    \phi_\mathrm{orb} = \mathrm{acos}(\vec{J}_\mathrm{orb} \cdot \vec{J}_\mathrm{gas,primary}). 
\end{equation}
Under this framework, a merger in which the secondary galaxy is perfectly aligned and co-rotating with the gas disk of the primary galaxy at infall has cos$\phi_\mathrm{orb} = 1$, while an aligned but counter-rotating secondary galaxy produces a merger with cos$\phi_\mathrm{orb} = -1$. A perpendicular merger would have cos$\phi_\mathrm{orb} = 0$.

In all three subsamples, we see a difference between the orientations of mergers that produce LSB galaxies and those that produce late-type HSB galaxies of similar mass. Both irregular and pure disk LSB galaxies are 4--5 times more likely than their corresponding late-type HSB samples to be the products of co-rotating mergers (defined here as $\mathrm{cos}\phi_\mathrm{orb}>$0.25), while bulge+disk LSB galaxies are twice as likely to have experienced this type of merger. We also find that LSB galaxies are 15--40\% less likely to have experienced a perpendicular merger (defined here as -0.25$\leq\mathrm{cos}\phi_\mathrm{orb}\leq$0.25  and indicated by dashed vertical lines in Figure \ref{fig:lsbmerg}) than their HSB counterparts, with the difference being most dramatic in the pure disk and bulge+disk groups. This suggests that \citet{dicintio2019}'s results do, indeed, hold at higher masses in \textsc{Romulus25}, but may be most relevant for galaxies with disky morphologies. A two-sample KS test rejects the hypothesis that pure disk or bulge+disk LSB galaxy merger orientations are drawn from the same distribution as those of HSB galaxies with p $<$ 10$^{-3}$, compared to p $\approx$ 0.01 for the irregular galaxy subsample. This difference in merger orientation also explains why the last major mergers of LSB galaxies produce more perturbation than those of HSB galaxies, as the timescales associated with aligned, co-rotating mergers are longer than those associated with perpendicular mergers and therefore contribute to a galaxy's PI over a longer period.

In the bottom panels of Figure \ref{fig:lsbmerg}, we show the distributions of orbital angular momenta for our classical LSB galaxies and mass-matched late-type HSB galaxies. Here, orbital angular momentum is calculated for the secondary galaxy at the time of infall. LSB galaxies within each of the subsamples tend to have experienced mergers with higher orbital angular momentum configurations than their HSB counterparts. This is most significant in the pure disk group, where the median orbital angular momentum is 0.53 dex higher among mergers that produce LSB galaxies, but also holds for the irregular and bulge+disk groups, where the median orbital angular momentum value among LSB galaxy progenitors is $\approx$0.4 dex higher than that of HSB galaxy progenitors. In all subsamples, a two-sample KS test rejects the hypothesis that the LSB and HSB galaxy orbital angular momenta are drawn from the same distribution with p $<$ $10^{-3}$. As orbital angular momentum is converted into the specific angular momentum of the remnant galaxy during the merger \citep[e.g.,][]{maller2002,vitvitska2002}, this---combined with more efficient momentum conversion via aligned, co-rotating mergers---is likely why we see more spin-up in LSB galaxies than in HSB galaxies following their respective final major mergers.

The idea that major mergers might be responsible for the creation of gas-rich, slowly evolving, typically disk-dominated objects like classical LSB galaxies is, in many ways, counter-intuitive. Traditionally, major mergers have been invoked as inciting transformations of galaxies from disky and star-forming to elliptical and quiescent. They are thought to drive gas to the centers of galaxies, possibly triggering AGN activity and causing galaxies to consume and eject their gas at a higher rate as they become more compact \citep[e.g.,][]{toomre1972,barnes1988,noguchi1988,hernquist1989,mihos1996,springel2005,Hopkins2008,stierwalt2013,Wilman2013}. While we do see a significant number of major mergers in \textsc{Romulus25} that follow this conventional path, these are predominantly the mergers that produce HSB galaxies. As may be seen in Figure \ref{fig:evorelmer2}, even those mergers that result in a late-type HSB galaxy, rather than an HSB elliptical, cause compactification and drive a decrease in spin and M$_\mathrm{HI}$/M$_\star$. However, it is clear that a second path exists: those mergers that produce LSB galaxies spin up both the dark and the baryonic matter, resulting in a gas-rich galaxy in which star formation is more spread out. This combination of gas-richness and inefficient star formation in turn leads to lower gas-phase metallicities and therefore less internal reddening due to dust, creating the appearance of a bluer disk. Major mergers, like galaxies, are diverse.

This challenge to the traditional view that mergers among massive galaxies always lead to quenching and disk destruction is neither new nor unique to the \textsc{NIHAO} and \textsc{Romulus25} simulations. \citet{Robertson2006} were the first to propose that high redshift mergers could contribute to the formation of rotationally supported disks and, building on earlier work from \citet{Barnes2002} and \citet{SpringelHernquist2005}, used idealized simulations to demonstrate that a variety of merger configurations could result in disks as long as the progenitors were gas-dominated and a sufficiently pressurized ISM model was used. Subsequent cosmological simulations have shown that gas-rich mergers can produce star-forming disk-dominated galaxies even at low redshifts \citep[e.g.,][]{Governato2009,Sparre2017,Jackson2020,Rodriguez2025}. In a study of more than 3000 galaxies in TNG300-1, \citet{Quai2021} find that, although post-merger galaxies are twice as likely to be quenched as control galaxies, merger-induced quenching is still rare, occurring in only $\sim$5\% of galaxies, most of which had a low gas fraction prior to the merger. Although \citet{Kulier2020} find that mergers typically cause galaxies to become more kinematically spheroid-dominated in the \textsc{EAGLE} simulations, they also note that the LSB galaxies with the most extended disks---comparable in size to observed giant LSB galaxies---are typically the products of major mergers that increased their spin. This is in keeping with a long history of both N-body and hydrodynamic simulations that have shown that mergers---both major and minor---are a possible route to the formation of giant LSB galaxies \citep[e.g.,][]{penarrubia2006,mapelli2008,zhu2018,saburova2018,Zhu2023}.

There is also some observational evidence that mergers may be related to the formation of LSB galaxies. Malin 1, the archetypal giant LSB galaxy, is currently interacting with at least two other galaxies \citep[][]{reshetnikov2010}, while compact dwarf galaxies thought to be the stripped remnants of more massive galaxies have been found on the outskirts of a number of other giant LSB galaxies \citep[e.g.,][]{Kasparova2014,hagen2016,saburova2018,Paswan2021,Saburova2021,Saburova2024}. At the other end of the mass spectrum, the disturbed morphologies of some LSB dwarf galaxies \citep[e.g.,][]{Egorova2021} and the abnormal globular cluster populations of some UDGs \citep[e.g.,][]{Silk2019,Saifollahi2022,Fielder2023} may suggest that mergers have played a significant role in their evolution. However, more direct signatures of mergers (e.g., stellar streams similar to those \citet{Galaz2015} discovered around Malin 1) tend to be extremely low surface brightness and to become less evident with time. We may need to rely on the next generation of telescopes to (in)validate our models of LSB galaxy formation.

\section{Conclusions}
\label{lsbsumm}

In this paper, we have identified a large sample of classical low surface brightness galaxies in the \textsc{Romulus25} cosmological simulation with the aim of comparing our galaxies to observed samples and determining how and why they differ from the more ``typical'' high surface brightness galaxy population. We find that these galaxies make up a significant fraction of the overall galaxy population, constituting 60\% of all central galaxies with 10$^8<$M$_\star$/M$_\odot<$10$^{10}$ (see Figure \ref{fig:frac}). In agreement with observations, our classical LSB galaxies tend to be extended (see Figures \ref{fig:struct} and \ref{fig:profs}), metal-poor (see Figure \ref{fig:methist}), and rich in neutral hydrogen (see Figure \ref{fig:HI}), but characterized by low star formation efficiencies (see Figure \ref{fig:SFRlsb}). However, we don't tend to reproduce the more extreme LSB galaxies that have been reported in observations, likely in part due to the relatively small volume of our simulation.

We find that the star formation histories of LSB galaxies are fairly similar to those of late-type HSB galaxies (see Figure \ref{fig:csfh}), but tend to be even steadier (i.e., characterized by a near-constant SFR) towards the low-mass end, leading to stellar populations that are, on average, slightly older and therefore slightly optically redder (see Figure \ref{fig:bvhist}). However, because LSB galaxies are diffuse and have relatively metal-poor gas---a feature that can be entirely accounted for by their HI-richness (see Figures \ref{fig:massmet} and \ref{fig:sbvmet})---they experience very little internal reddening and may therefore appear bluer than HSB galaxies of similar stellar mass (see Figure \ref{fig:bvdust}).

The major differences between late-type LSB and HSB galaxies---HI-richness (and therefore metal deficiency and blueness), spatially extended, inefficient star formation, and LSB centers---appear to be the products of major mergers (see Figure \ref{fig:evorelmer2}). Like the field ultra-diffuse galaxy samples studied in \citet{wright2021} and \citet{VanNest2022} and the small sample of LSB galaxies analyzed in \citet{dicintio2019}, classical LSB galaxies in \textsc{Romulus25} are typically significantly spun up by their last major merger. In mergers that produce LSB galaxies, the secondary galaxy is typically co-rotating and aligned with the gas disk of the primary galaxy and/or has higher orbital angular momentum at infall, allowing for a larger amount of orbital angular momentum to be converted into the internal angular momentum of the resultant galaxy (see Figure \ref{fig:lsbmerg}). By contrast, major mergers that produce HSB galaxies are more likely to be perpendicular and typically result in a decrease in spin. This likely contributes to the more compact distribution of stars and star formation in HSB galaxies, leading to more efficient star formation and higher gas consumption rates---the traits that we typically associate with major mergers among massive galaxies.

LSB galaxies tend to remain relatively spun up long after their mergers have concluded. Consequently, their gas and therefore the stars that ultimately form out of that gas tend to be increasingly spatially extended and diffuse, particularly among higher-mass galaxies. Consequently, star formation rates within the centers of the galaxies are low, resulting in lower central surface brightnesses and shallow surface brightness profiles. In LSB galaxies with bulges, these lower central star formation rates produce older, redder bulges than we see in comparably massive HSB galaxies with bulges. Although these extended, gas-rich galaxies do not fit the traditional expectation for major merger products, they represent an equally valid---and not terribly uncommon---path in galaxy evolution.

This is a crucial time for studies of the LSB universe. Recent advances in imaging and post-processing techniques have led to a boom in large-scale LSB galaxy surveys \citep[e.g.,][]{greco2018illuminating,Zaritsky2019,Danieli2020,Tanoglidis2021,Prole2021,Kado-Fong2021,Zaritsky2022,Greene2022}. Over the next decade, a number of instruments designed with low surface brightness science in mind will be seeing first light, including the \textit{Nancy Grace Roman Space Telescope} \citep[e.g.,][]{Montes2023}, the Vera Rubin Observatory \citep[e.g.,][]{Martin2022}, and \textit{Euclid} \citep[e.g.,][]{EuclidCollaboration2022}, the latter two of which are already taking data. We might therefore expect some of the most exciting discoveries of the late 2020s and 2030s to be made in the low surface brightness universe. The role of simulations will be to help interpret these discoveries and to aid in predicting where the next discovery might be made. However, for that to be possible, simulations must be capable of reproducing and explaining current observations of the low surface brightness universe. To that end, we will continue to explore how the detailed merger and tidal histories of LSB galaxies influence their evolution in future work (Corillo in prep.).

\section*{Acknowledgments}
The authors thank the anonymous referee for their thoughtful review, which improved the content and clarity of the paper. The authors also thank Julianne Dalcanton, Shany Danieli, Arianna Di Cintio, Marla Geha, Johnny Greco, Jenny Greene, Daisuke Nagai, Laura Sommovigo, and Stephanie Tonnesen for useful discussions related to this work. ACW is supported by an ACM SIGHPC/Intel Computational \& Data Science fellowship and the \textit{Nancy Grace Roman Space Telescope} Project, under the Milky Way Science Investigation Team. AMB acknowledges support by grant FI-CCA-Research-00011826 from the Simons Foundation. FDM acknowledges support by NSF AAG-2407232 and Phy-2013909. The python packages {\sc matplotlib} \citep{hunter2007}, {\sc numpy} \citep{walt2011numpy}, {\sc tangos} \citep{pontzen2018}, {\sc pynbody} \citep{pynbody}, \textsc{scipy} \citep{scipy2020}, {\sc Astropy} \citep{astropy2013,astropy2018}, and \textsc{seaborn} \citep{Waskom2021} were all used in parts of this analysis. This research is part of the Blue Waters sustained-petascale computing project, which is supported by the National Science Foundation (awards OCI-0725070 and ACI-1238993) and the state of Illinois. Blue Waters is a joint effort of the University of Illinois at Urbana-Champaign and its National Center for Supercomputing Applications. This work is also part of a PRAC allocation support by the National Science Foundation (award number OAC-1613674).  Initial condition generation and some analysis was performed on ACCESS (formerly XSEDE) resources at SDSC and TACC.  ACCESS is supported by National Science Foundation grants 2138259, 2138286, 2138307, 2137603, and 2138296.
 \section*{Data Availability}
The data for this work were generated from a proprietary branch of the {\sc ChaNGa} N-Body+SPH code \citep{menon2015}. The public repository for {\sc ChaNGa} is available on github (https://github.com/N-BodyShop/changa). Analysis was conducted using the publicly available softwares pynbody \citep[][https://github.com/pynbody/pynbody]{pynbody} and TANGOS \citep[][https://github.com/pynbody/tangos]{pontzen2018}. These results were generated from the \textsc{Romulus25} cosmological simulation. The raw output from this simulation can be accessed upon request from Michael Tremmel (mtremmel@ucc.ie), along with the TANGOS database files that were generated from these outputs and directly used for this analysis.
\bibliographystyle{aasjournalv7}
\bibliography{LSBs}
\label{lastpage}
\end{document}